\documentclass[reprint,aps,pra,twocolumn,superscriptaddress,amsmath,amssymb]{revtex4-1}
\usepackage{graphicx}  % needed for figures
\usepackage{dcolumn}   % needed for some tables
\usepackage{bm}        % for math
\usepackage{verbatim}   % for math
\usepackage[mathscr]{euscript}
\usepackage{stmaryrd}
\usepackage{centernot}
\usepackage{slashed}
\newcommand{\lbar}{\mathcal{I}}

\newcommand{\beq}{\begin{equation}}
\newcommand{\eeq}{\end{equation}}
\newcommand{\beqa}{\begin{eqnarray}}
\newcommand{\eeqan}{\end{eqnarray*}}
\newcommand{\beqan}{\begin{eqnarray*}}
\newcommand{\eeqa}{\end{eqnarray}}
\newcommand{\bea}{\begin{eqnarray}}
\newcommand{\eea}{\end{eqnarray}}
\newcommand{\ket}[1]{\left| #1 \right\rangle}
\newcommand{\bra}[1]{\left\langle #1 \right|}
\newcommand{\eq}[1]{Eq.~(\ref{#1})}
\newcommand{\eqs}[1]{Eqs.~(\ref{#1})}
\newcommand{\eqr}[1]{(\ref{#1})}
\bibpunct{[}{]}{,\kern-0.3em}{n}{}{}   %changes superscript citations to online ones, e.g [2]
\newcommand{\up}{\uparrow}
\newcommand{\dn}{\downarrow}
\newcommand{\ip}[1]{\langle{#1}\rangle}
\newcommand{\tr}{{\mathrm{Tr}}}

\usepackage{color,xcolor}
\usepackage[normalem]{ulem}
\newcommand\Note[2]{\textcolor{cyan}{[[ Note - #1 ]] {#2} [[ End note ]]}}

\ifdefined\shortcomments

  \renewcommand\Note[2]{}
  
\fi

\begin{document}

\title{Quantum heat engine operating between thermal and spin reservoirs}
\author{Jackson S. S. T. Wright}
\affiliation{
   Centre for Quantum Dynamics,\\
   Griffith University,\\
   Brisbane, QLD 4111 Australia
   }
\author{Tim Gould}
\affiliation{
  Queensland Micro- and Nanotechnology Centre,\\
   Griffith University,\\
   Brisbane, QLD 4111 Australia
   }
\author{Andr\'e R. R. Carvalho}
\author{Salil Bedkihal}
\author{Joan A. Vaccaro}
\affiliation{
   Centre for Quantum Dynamics,\\
   Griffith University,\\
   Brisbane, QLD 4111 Australia
   }

\date{\today}
\begin{abstract}
Landauer's erasure principle is a cornerstone of thermodynamics and information theory.
According to this principle, erasing information incurs a minimum energy cost.
Recently, Vaccaro and Barnett [Proc. R. Soc {\bf 467}, 1770 (2011)] explored information erasure in the context of multiple conserved quantities and showed that the erasure cost can be solely in terms of spin angular momentum.
As Landauer's erasure principle plays a fundamental role in heat engines, their result considerably widens the possible configurations that heat engines can have.
Motivated by this, we propose here a novel optical heat engine that operates under a single thermal reservoir and a spin angular momentum reservoir coupled to a three level system with an energy-degenerate ground state. The proposed heat engine operates without producing waste heat and goes beyond the traditional Carnot engine where the working fluid is subjected to two thermal baths at different temperatures.

\end{abstract}

\maketitle

\section{Introduction}
Maxwell's demon, a hypothetical intelligent being that appears to violate the second law of thermodynamics, has been widely studied since it was proposed in 1867 \cite{Maxwell}.
In the original formulation of this thought experiment, the being controls a small hole in a division between two portions of a vessel that contains a gas.
The being opens and shuts the hole as individual gas molecules approach it so that fast molecules pass through the hole in one direction and slow ones in the opposite direction leading to one portion of the gas warming up and the other cooling down.
The demon can then extract work from the gas using a Carnot heat engine.
This process contradicts the Kelvin-Planck statement of the second law of thermodynamics that it is impossible to extract a net amount of work from a single thermal reservoir.

Bennett \cite{Bennett1982} proposed resolution to the paradox lay in recognizing that the memory of the demon is changed in the process of determining the speeds of the molecules, and to complete a full thermodynamic cycle the memory needs to be reset to its initial state.
According to Landauer’s principle, any logically irreversible transformation of classical information is necessarily accompanied by the dissipation of at least $k_B T\ln(2)$ of heat per bit, where $k_B$ is the Boltzmann constant and $T$ is the temperature of the reservoir that absorbs the heat \cite{Landauer1961}.
Bennett gave a demonstration where the cost of erasing the demon's memory is at least as much as the work the demon has extracted, and so the validity of the Kelvin-Planck statement of the second law is restored.

%The idea of heat engines based on the working fluid interacting sequentially with a stream of systems or reservoirs, has recently received attention \cite{Mandal2012},\cite{Strasberg2017}. The Maxwell's demon in these cases operates autonomously, without external control.
The role of Maxwell's demon in the operation of heat engines has become more transparent over time.  For example, Mandal and Jarzynski \cite{Mandal2012} and Strasberg \emph{et al}. \cite{Strasberg2017} have recently proposed heat engines in which the working fluid interacts sequentially with a stream of systems or reservoirs.  The Maxwellian demon in these models operates autonomously, without external control, and allows a complete analysis.
Recent technological advances in nanotechnology and photonics have
allowed their experimental realizations.  For example, an electronic version of Maxwell's demon
using two coupled single electron transistors %devices (SET) has been
realized by Koski \emph{et al}. \cite{Pekola2015} in an experiment
where each SET played the role of the demon or system that the demon
interacts with.  A photonic version of Maxwell's demon has been
realized by Vidrighin \emph{et al}. \cite{Vidrighin2016} in an
experiment that extracts work from thermal light. In these
realizations the erasure of the demon's memory occurs at a cost in
terms of energy.

The association between information erasure and energy embodied in
Landauer's bound is widely accepted as a natural one in traditional
thermodynamics.  However, in two classic papers, Jaynes
\cite{Jaynes1957, Jaynes1957b} formulated a generalized theory of
statistical mechanics using the {\it maximum entropy principle} where
not only energy but all other {\it measurable conserved} quantities
can be treated on an equal footing.  In this framework the notion of
heat can be generalized to incorporate an exchange of arbitrary
conserved quantities such as quantized spin angular momentum. In other
words if there are $N_k$ conserved quantities associated
with the generalized reservoir, then for the conserved quantity
labelled $k$ the corresponding heat is called the ``$k^{th}$'' heat
and the corresponding measurement probe is called the
``$k^{th}$'' meter \cite{Jaynes1957, Jaynes1957b}. If
the ``$k^{th}$'' conserved quantity corresponds to energy
then the corresponding meter is the thermometer.

In 2006 Vaccaro and Barnett \cite{Vaccaro2006,Vaccaro2011, Barnett2013,Croucher2018} applied Jaynes' maximum entropy framework to the problem of erasing information when multiple conserved quantities are present.
Using the generalized Gibbs ensemble they argued that the cost of erasure can be in terms of a ``$k^{th}$" conserved quantity.
In particular, they described an information-erasure model based on an energy-degenerate spin reservoir for which the cost of erasure, $\mathcal{L}_{s}$, is solely in terms of the dissipation of a minimum amount of spin angular momentum given by
\bea
   \mathcal{L}_{s} \geq
    \gamma^{-1}\ln{2} \label{eqn:VB}
\eea
per bit, where
\bea
    \gamma=\frac{1}{\hbar}\ln\left[\frac{N\hbar-2\langle
    \hat{J}^{(R)}_{z}\rangle}{N\hbar+2\langle
    \hat{J}^{(R)}_{z}\rangle}\right]=\frac{1}{\hbar}\ln\left[\frac{1-\alpha}{\alpha}\right]\label{eqn:gamma}
\eea
is an inverse spin temperature and $\alpha$ denotes the degree of spin polarization of the spin reservoir.
The quantity $\mathcal{L}_{s}$ is called the \emph{spinlabor} \cite{Toshio2017}.
It is the spin equivalent of work and is defined as the amount of spin angular momentum that is supplied by an external source and transferred to (i.e.
dissipated in) the spin reservoir during the erasure process.
In analogy with the transfer of heat to a thermal reservoir, the change in the spin angular momentum of the spin reservoir is called \emph{spintherm} $\mathcal{Q}_{s}$ \cite{Toshio2017}.
Croucher \emph{et al.} have shown that $\mathcal{L}_{s}$ is an average over many erasure processes and that the cost for a single erasure process has discrete fluctuations that satisfy a Jarzynski-like equality \cite{Toshio2017}.
If the energy degeneracy of the spins is broken by a Zeeman field then the erasure will incur costs in terms of both energy and spin polarization \cite{Vaccaro2011, Barnett2013}.

This new method of erasure allows novel kinds of heat engines.  For example, Vaccaro and Barnett \cite{Vaccaro2011} proposed a heat engine that uses a Maxwell demon to extract work from a single thermal reservoir and an energy-degenerate spin reservoir to erase the memory of the demon each cycle.  In contrast to Bennett's resolution \cite{Bennett1982} of the Maxwell demon paradox in which the demon produces no net work output, the proposed spin-heat engine (SHE) gives a net output of work for an erasure cost in angular momentum.

Following Vaccaro and Barnett's work \cite{Vaccaro2011,Barnett2013}, a number of groups have explored thermodynamical resource theories in which there are multiple conserved quantities and where the conventional thermal reservoir is replaced with one satisfying a generalized Gibbs ensemble.
For example, Guryanova \emph{et al.} \cite{PST11} showed that each conserved quantity can be extracted from a generalized reservoir provided that other conserved quantities are supplied.  In other words, a generalized Gibbs reservoir can be used to exchange one conserved quantity for another.  Their result generalizes and extends the exchange of angular momentum for work that would operate in the heat engine proposed by Vaccaro and Barnett \cite{Vaccaro2011}.
A heat engine that exchanges spin angular momentum for an induced
electrical current in a quantum spin Hall device has recently been
proposed by Bozkurt \emph{et al.}  \cite{Bozkurt2017}.
Yunger {}Halpern \emph{et al.} introduced an approximate
microcanonical ensemble to deal with non-commuting conserved
quantities \cite{Halpern2015}.
Lostaglio \emph{et al.} have explored  tradeoffs between costs of
information erasure in the case of multiple conserved quantities
extending the work of Vaccaro and Barnett \cite{Barnett2013}, and also
highlighted difficulties associated with non-commutativity
\cite{Lostaglio2017}.

In this paper we describe in more detail the conceptual basis of the
SHE proposed by Vaccaro and Barnett \cite{Vaccaro2011}. This will be the first analysis of a quantum dot heat engine that is coupled to both thermal and nuclear spin reservoirs. We first formalize the general operation of the SHE and give an example in terms of an archetypical optical heat engine in
section II. In section III we describe the theory underpinning a
possible realization in quantum dot technology and in section IV we
present results of simulations. We end with a discussion and
conclusion in section V. Technical details are left to an
Appendix.

\section{Conceptual basis}
\label{Sec:conceptual}
\begin{figure}[b]
\centering %\captionsetup{justification=RaggedRight}
  \includegraphics[width=0.40\textwidth]{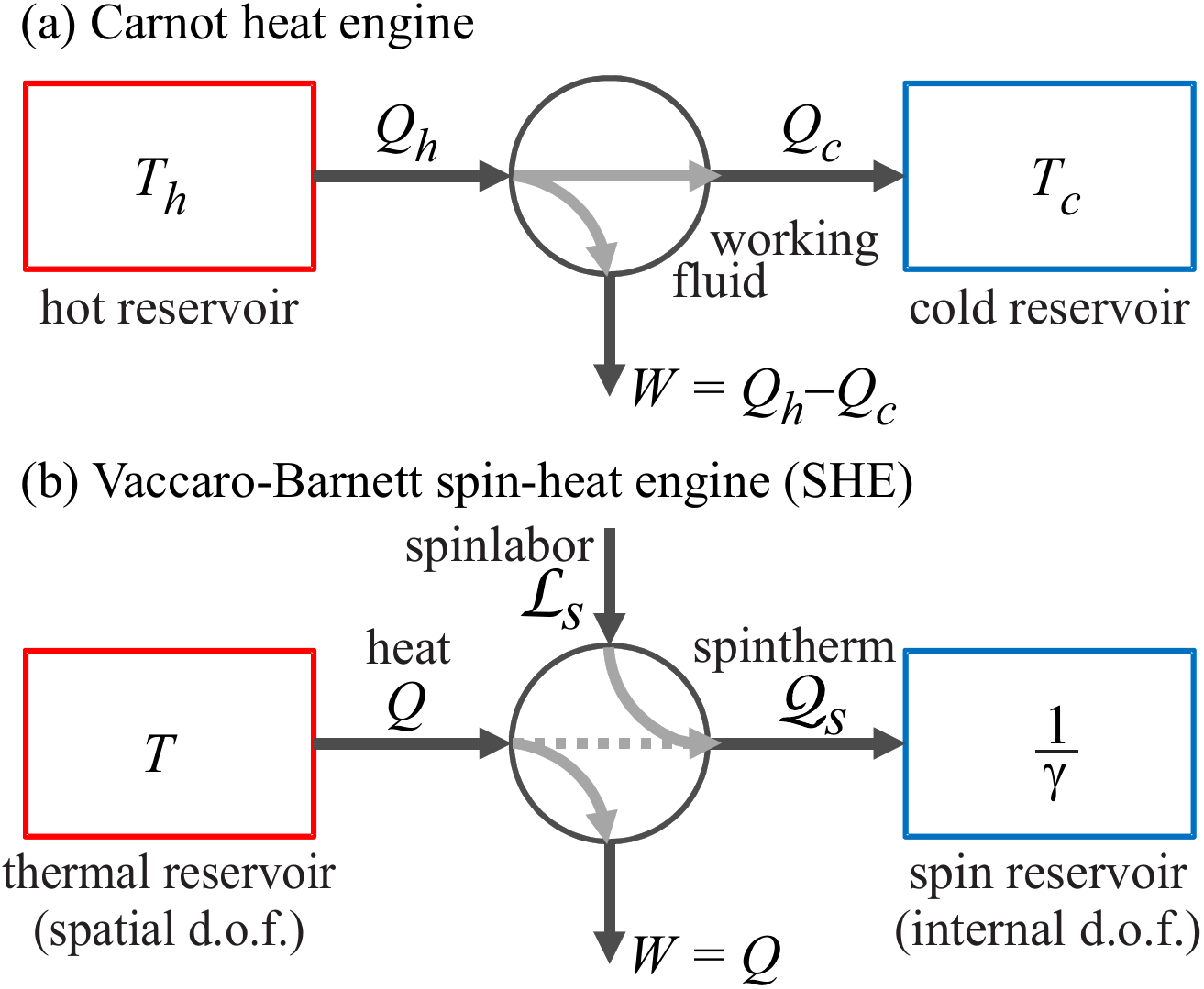}
\caption{Conceptual diagrams for (a) a conventional Carnot heat engine
  and (b) the Vaccaro-Barnett spin-heat engine that operates between a thermal
  reservoir and a spin reservoir. In (b) the thermal reservoir would
  typically be represented by a spatial degree of freedom and the spin
  reservoir by an internal degree of freedom, and $\gamma$ is the
  inverse spin temperature associated with the spin reservoir.}
  \label{fig: comparison of Carnot engine with a SHE}
\end{figure}

The general operation of the SHE is best explained by comparing it to the conventional Carnot heat engine.
Unlike the Carnot heat engine,
Fig.~\ref{fig: comparison of Carnot engine with a SHE}(a), which
operates between two thermal reservoirs at different temperatures, the
SHE, Fig.~\ref{fig: comparison of Carnot engine with a SHE}(b),
operates between a single thermal reservoir and a spin reservoir.
In the SHE, a quantity of heat $Q_{h}$ is extracted from the thermal
reservoir and converted entirely to optical work $W=Q_{h}$ which
represents an efficiency of $\eta=W/Q_{h}=1$.
The conversion results in lower entropy and so, by the second law, a
compensatory increase in entropy must appear elsewhere.
In contrast to the case for the Carnot engine where the increase
occurs in the cooler thermal reservoir via the delivery of waste heat,
in the SHE the
\emph{increase occurs as a result of information erasure}
which is accompanied by dissipating an amount of spinlabor
$\mathcal{L}_s$ as spintherm $\mathcal{Q}_s$ in the spin reservoir.
Essentially, the entropy is decoupled from the energy that is
extracted from the thermal reservoir and transferred to the spin
reservoir as spintherm $\mathcal{Q}_s$.
The net effect is that the thermal reservoir cools down, the spin
reservoir gains entropy which reduces its polarization, and work is
extracted. In the situation where one considers the execution of multiple engine cycles, then a careful accounting of the entropy change in the reservoir becomes important. Such calculation would show how many cycles the finite nuclear spin reservoir would sustain before the need to be recharged (reset to its initial fully polarised state). Here, we will consider that the spin bath is large enough so that a few cycles can be performed without the need to a reservoir reset.

In the remainder of the paper we will present an analysis for a potential realization of this SHE in quantum dot
(QD) systems. Note however that this archetypical SHE may also be realized in a number of different technologies, e.g. semiconductor heterostructures, spin
polarized gases, ion-traps and hybrid optomechanical systems. In fact, a recent example is the model of a quantum spin Hall heat engine proposed by Bozkurt {\em{et al}} \cite{Bozkurt2017}. Although not previously noticed by the authors, their model operates according to the same principles as the SHE \cite{Vaccaro2011} and so we can use it to illustrate the conceptual basis of the SHE.
In their model, the electrons represent the thermal reservoir, and the nuclear spins in the vicinity of the edge spin currents represent spin reservoirs.
The nuclear spin reservoir induces electrons to undergo a spin flip and backscatter from one spin current into the spin current flowing in the opposite direction.
For definiteness, consider spins flipping from up to down (alternatively, down to up) with respect to an appropriate spin direction and let the change in chemical potential be $\Delta\epsilon$.
For each electron taking part in this process, the nuclear spin reservoir absorbs spintherm of $\mathcal{Q}_s=1\hbar~(-1\hbar)$, the thermal reservoir supplies the energy needed as heat  $Q=\Delta\epsilon$, and work of $W=\Delta\epsilon$ and spinlabor of $\mathcal{L}_s=-1\hbar~(1\hbar)$ are done on the spin currents.

\section{Quantum dot spin-heat engine}

\subsection{The system}

\begin{figure}[t]
\centering %\captionsetup{justification=RaggedRight}
  \includegraphics[width=0.45\textwidth]{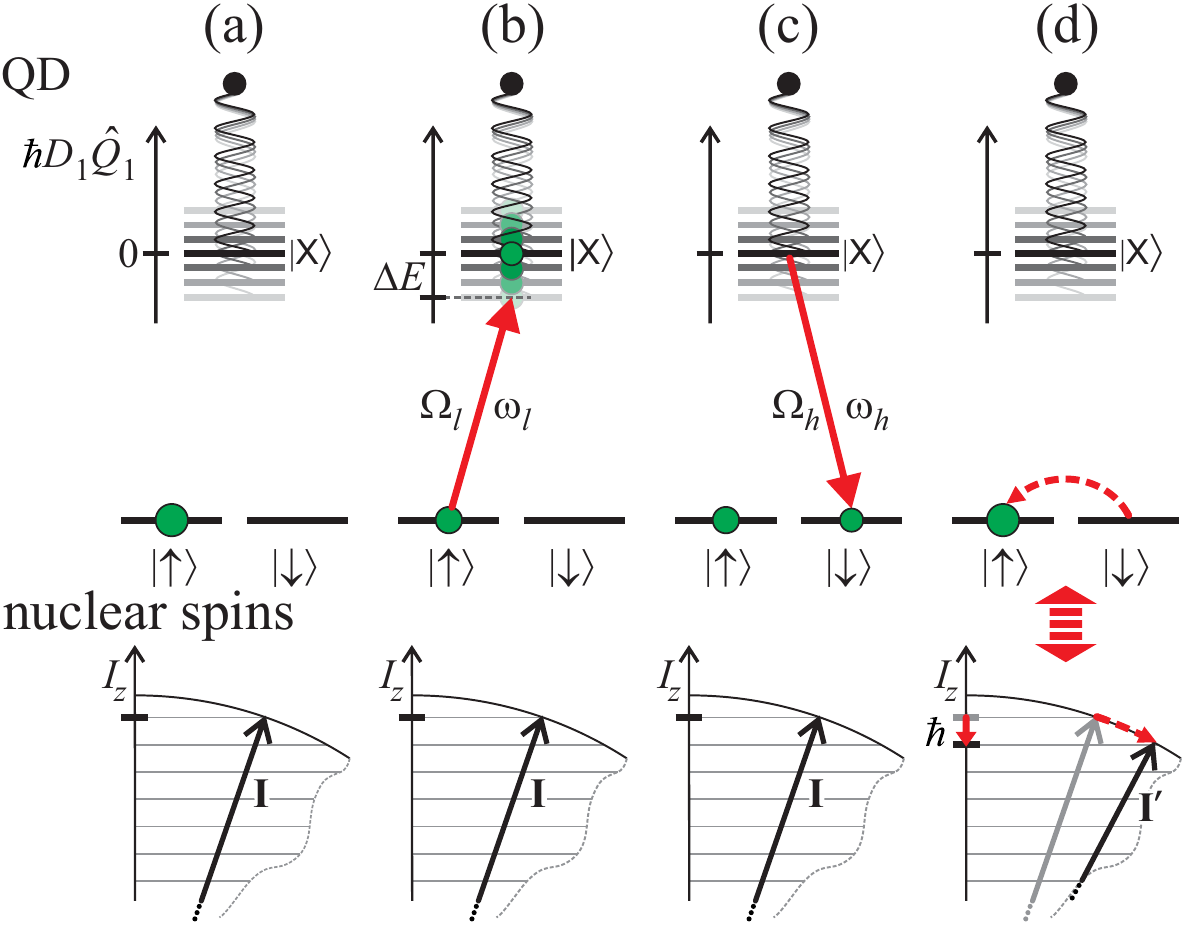}
  \caption{Stages of the qdSHE.
  The interaction between the QD and the phonons is represented by a dynamic shift in the energy of the exciton state. The nuclear spins, which are initially polarized in the direction of the $z$ axis, are represented by their total spin angular momentum $\mathbf{I}$ and its projection onto the $z$ axis.
  Panel (a) represents the beginning of a cycle, (b) represents the extraction of $\Delta E$ of heat by a red detuned laser, (c) represents the conversion of the extracted heat to $\Delta E$ of optical work by an on-resonance laser, and (d) represents the spin exchange between the QD and the nuclei that erases information left in the QD.
  The size of the green disc symbolizes the relative population in each state of the working fluid.
    \label{fig:QD stages}}
\end{figure}

\paragraph*{Working fluid.}
Three electronic states, viz. $\ket{\up}$, $\ket{\dn}$ and $\ket{X}$, of the QD compose the working fluid of our QD spin-heat engine (qdSHE).
Their energy level structure is depicted in Fig.~\ref{fig:QD stages}.
The states $\ket{\up}$, $\ket{\dn}$ represent ground-level spin states and $\ket{X}$ is an excited state that has optically allowed transitions from the ground states.
For example, in a negatively-charged self-assembled QD, the ground states would be $\ket{\up}=\hat{e}^\dagger_{\up}\ket{0}$ and $\ket{\dn}=\hat{e}^\dagger_{\dn}\ket{0}$ and the excited state would be an exciton such as $\ket{X}=\hat{e}^\dagger_{\dn}\hat{e}^\dagger_{\dn}\hat{h}^\dagger_{3/2}\ket{0}$, where $\ket{0}$ represents the crystal ground state and $\hat{e}^\dagger_{\mu}$, $\hat{h}^\dagger_{\mu}$ are electron and hole creation operations associated with spin state $\mu$.

\paragraph*{Thermal reservoir.} The phonons of the supporting crystal play the role of the thermal reservoir in our qdSHE.
For longitudinal acoustic phonons, which are the main source of dephasing in self-assembled quantum dots, the interaction between the QD and the phonons is a spin-boson type \cite{Mahan2000} and is represented in Fig.~\ref{fig:QD stages} by a shift of $\hbar D_{1}\hat Q_1$ in the energy of the exciton state $\ket{X}$ where $\hat Q_1$ is the position operator of an effective phonon mode and $D_{1}$ is a coupling constant; the details of this interaction are explained in section \ref{sec:QD-phonon interaction}.
The phonons are assumed to be in a thermal state initially, and so the initial mean and variance of $\hat Q_1$ are $\ip{\hat{Q}_1}=0$ and $\ip{\Delta\hat{Q}^2_1}\ne0$.
However, laser excitation of the QD induces oscillation in $\hat Q_1$, and these oscillations manifest non-Markovian behavior of the phonons over times of order 1 ps.
 The intense peaks in a photoluminescence spectrum of this system correspond to the energy of the exciton at $\hat Q_1=0$ and are called the zero phonon lines.

\paragraph*{Spin reservoir.}In a negatively-charged QD, the conduction band electron has a wave function that extends over $10^3$ to $10^5$ nuclei \cite{Imamoglu2003}.
These nuclei constitute the spin reservoir for the qdSHE.
The experimental manipulation of the nuclear spins using conduction band electrons date back to the 1960's \cite{Lampel1968}.
A technique for polarizing the nuclear spins, known as dynamic nuclear polarization (DNP), has been proposed for QDs \cite{Imamoglu2003,Christ2007} with the degree of polarization predicted to be up to $99$\%.
In 2010 Chekhovich {\it et al.} \cite{Chekhovich2010} achieved $65$\% polarization of the nuclear spins experimentally.
We assume that a technique like DNP has been carried out prior to operating the qdSHE and that the ideal case of near perfect nuclear spin polarization in the $z$ direction has been reached.
The nuclear spin reservoir is represented in the lower portion of each panel of Fig.~\ref{fig:QD stages} by the mean total angular momentum $\mathbf{I}$ and its projection onto the $z$ axis.

\subsection{Operating cycle of the qdSHE}

The spin-boson character of the phonon-QD interaction, which is represented in Fig.~\ref{fig:QD stages} by the effective mode causing a fluctuating shift in the energy of the exciton $\ket{X}$, means that the phonons do work on the QD during the times when the energy of the exciton is being raised.
The qdSHE exploits this property by purposely driving population into the exciton when its energy is lower than average, and driving it out of the exciton when its energy has been raised.
This results in a three-stage cycle consisting of a heat extraction stage, a work output stage and an erasure (or resetting) stage.
A cycle of operation of the qdSHE begins with the QD in the ground state $\ket{\up}$ as depicted in Fig.~\ref{fig:QD stages}(a)

\paragraph*{Heat extraction.}
This stage is designed to allow the phonons to do work on the QD, on average.
A laser that is red-detuned by $\Delta E$ from the zero phonon line of the $\ket{\up}\leftrightarrow\ket{X}$ transition, as depicted in Fig.~\ref{fig:QD stages}(b), is directed at the QD.
It induces significant population transfer only when the fluctuating shift in the energy of $\ket{X}$ compensates for the detuning.
Any population that is transferred to $\ket{X}$ is subject to the fluctuating energy shifts induced by the effective phonon mode.  The effective phonon mode does $\Delta E$ of work on the transferred population, on average, in raising its energy to the time-averaged value of the state $\ket{X}$.

\paragraph*{Optical work output.}  For this stage, a laser that is resonant with the zero phonon line of the transition $\ket{\dn}\leftrightarrow\ket{X}$, as depicted in Fig.~\ref{fig:QD stages}(c), transfers population from $\ket{X}$ to $\ket{\dn}$.
The duration of the stage is timed so that as much of the population in $\ket{X}$ is transferred.
The work of $\Delta E$ appears as increased coherent light.

\paragraph*{Erasure (or resetting).}
The spin-exchanging hyperfine interaction between the nuclei and the electron of the QD is ever present.
However, it occurs on a time scale of nanoseconds which is much slower than the tens of picoseconds time scale of the laser pulses and picosecond time scale of the phonons and so its effect is negligible during the previous stages.
The erasure stage, represented in Fig.~\ref{fig:QD stages}(d), consists simply of waiting for a duration of the order of tens of nanoseconds to allow the hyperfine interaction to bring the spin of the electron into  equilibrium with the spins of the nuclei.

%The complete cycle of the qdSHE is described in terms of the heat-work-spinlabor-spintherm flow diagram in Fig.~\ref{fig: comparison of Carnot engine with a SHE}(b) as follows.
%For each $\omega_l$ photon absorbed in the heat extraction stage, one $\omega_h$ photon is emitted in the work output stage, giving a net increase of $W=\hbar(\omega_h-\omega_l)=\Delta E$ in coherent light. This optical work is the result of $Q=\Delta E$ of heat being extracted from the phonons as they do work on the QD.
%The price paid for this is the expenditure of $\mathcal{L}_s=1\hbar$ of spinlabor manifested as a net loss of $1\hbar$ of spin angular momentum from the combined laser fields as a result of inducing the transition from $\ket{\up}$ to $\ket{\dn}$ in the working fluid.
%This amount of spinlabor is dissipated as spintherm $\mathcal{Q}_s=\mathcal{L}_s$ in the spin reservoir during the erasure stage.

In the remainder of this section we describe the phonon and nuclear spin interactions with the QD.

\subsection{Modelling the QD-phonon interaction \label{sec:QD-phonon interaction}}

There have been a number of studies of the role of phonons in the photoluminescence induced in QDs, both in the presence of an optical cavity or otherwise
\cite{Besombes2001,Wilson-Rae2002,Roy2011a,McCutcheon2013,Roy2011b,
Timm2011,Stock2011,Majumdar2012,Kaer2012,Weiler2012,McCutcheon2013,Ulhaq2013}.
The interaction between the longitudinal acoustic phonon modes and the electronic system is described by a spin-boson coupling whose strength depends on the electronic  spatial wave functions \cite{Mahan2000}.
The ground electronic states have the same orbital wave function and so they couple equally to the phonon modes.
Provided the electron lies predominately in the ground states for a time longer than the coherence time of the phonon modes, the phonon modes will find an equilibrium that incorporates the deformation due to the ground states; the details of this argument can be found, e.g., in appendix B of \cite{Kaer2012} and also section III of \cite{Roszak2005}.
We assume that this is the case here, and so essentially the phonon coupling to the electronic states can be incorporated in a modified coupling to the exciton state alone.

As shown in Fig.~\ref{fig:QD stages}, at various times a coherent laser drives the transition between one of the ground states and the exciton state.
The corresponding Hamiltonian describing the electronic system, phonon bath and a laser field system is given, in the rotating frame with respect to the laser frequency and after making the rotating wave approximation, by %[NOTE $\hbar\lambda_k$ - SEE [19]. REF [34] USES $\lambda_k$]
\begin{align}
    \label{eq:Ham momentum rep}
  \hat{H}_{\rm ep} &=\hbar\Omega_\mu(\hat{\sigma}^{+}_\mu+\hat{\sigma}^{-}_\mu)
  +\ket{X}\bra{X}\left[\hbar\Delta_\mu+\sum_{k}\hbar\lambda_{k}(\hat b^{\dagger}_{k}+\hat b_{k})\right]\nonumber
  \\&\qquad+\sum_{k}\hbar\omega_{k}\hat b^{\dagger}_{k}\hat b_{k},
\end{align}
where $\Delta_\mu$, $\Omega_{\mu}$ and $\sigma^{+}_\mu=\ket{X}\bra{\mu}=(\hat{\sigma}^{-}_\mu)^\dagger$ are the detuning of the laser, Rabi frequency and raising operator associated with the laser-driven transition $\ket{\mu}\leftrightarrow\ket{X}$ for $\mu=\up$ or $\dn$, $\lambda_{k}$ is the modified phonon coupling parameter for the exciton state, and $\hat{b}_k^\dagger$, $\hat{b}_k$ are the creation and annihilation operators of the phonon mode with momentum indexed by $k$.
\eq{eq:Ham momentum rep} shows that the coupling between the phonon modes and the QD results in the energy of the exciton state being shifted by a weighted sum of phonon position operators $(\hat b_k^\dagger+\hat b_k)/\sqrt{2}$.
As the position is inherently oscillating, the energy of the exciton state $\ket{X}$ is represented in Fig.~\ref{fig:QD stages} as being attached to the end of a spring.

The correlation time of the phonons is of the order of picoseconds which is sufficiently long for the correlations between them and the quantum dot to be important for the operation of the qdSHE.
A non-Markovian treatment of the phonons is therefore needed.
There are many ways this can be done.
For example, non-Markovian master equations can be based on perturbative expansions of a memory kernel \cite{machnikowski2004}.
Microscopic methods to simulate non-Markovian dynamics include non-equilibrium Green's functions \cite{Wilner2014}, numerically exact deterministic iterative path integral schemes \cite{Weiss2013}, and path integral quantum Monte-Carlo methods \cite{Mason1989}.
However, these approaches are often numerically expensive to implement.
The chain representation of open quantum systems offers an alternate technique for simulating the short time quantum dynamics accurately with relative computational ease \cite{Hughes2009a}.
The philosophy behind it is similar to the time dependent density matrix renormalization group methods where the bath is discretized and the spectral density is evaluated recursively until the desired convergence is obtained.
The key advantage of the chain representation is that the dynamics can be often accurately obtained using a truncated Hilbert space \cite{Hughes2009a}.

Here we adopt the truncated chain representation developed by Burghardt and coworkers \cite{Burghardt2005,Burghardt2006a,Burghardt2006b,Burghardt2007,Gindensperger2007,
Burghardt2008a,Burghardt2008b,Burghardt2008b,Hughes2009a,Hughes2009b} to model the non-Markovian QD-phonon interaction.
In this approach, the weighted sum of phonon position operators appearing in \eq{eq:Ham momentum rep} is identified as the position operator $\hat Q_1$ of an \emph{effective mode} where
\begin{align}
        \hat{Q}_{1} &=\frac{1}{D_1}\sum_{k}\lambda_{k}(b^{\dagger}_{k}+b_{k})\ ,\\
        \label{eq:D definition}
        D_1^2&=\sum_{k}|\lambda_{k}|^{2}\ .
\end{align}
The momentum operator and angular frequency associated with the mode are given by
\begin{align}
        \hat{P}_{1}&=\frac{i\hbar}{2D_1}\sum_{k}\lambda_{k}(b^{\dagger}_{k}-b_{k})\ , \\
        \label{eq:tilde omega definition}
        \tilde\omega_1^2&=\frac{1}{D_1^2}\sum_{k}\omega_k^2|\lambda_{k}|^{2}
\end{align}
where $[\hat{Q}_{1},\hat{P}_{1}]=i\hbar$.
This effective mode is used to define an infinite chain of orthogonal effective modes that comprise different linear combinations of the momentum modes.
The Hamiltonian in \eq{eq:Ham momentum rep} then takes the following form (in mass-weighted coordinates \cite{Hughes2009a})
\begin{align}
    \label{eq:Hamiltonian full chain}
   \hat H_{\rm ep} &=\hbar\Omega_\mu(\hat{\sigma}^{+}_\mu+\hat{\sigma}^{-}_\mu)
   +\ket{X}\bra{X}(\hbar\Delta_\mu+\hbar D_1\hat{Q}_{1})\nonumber\\
   &+\frac{1}{2}\sum_{n}[\hat{P}_{n}^2+\widetilde\omega_{n}^2\hat{Q}_{n}^2]\ ,
\end{align}
where the last term represents the phonon bath Hamiltonian in terms of the position and momentum operators $\hat{Q}_{n}$ and $\hat{P}_{n}$, and frequencies $\widetilde\omega_{n}$, of the effective modes \cite{Gindensperger2006}.

In our case, it is sufficient to retain only the most significant non-Markovian effects and these are obtained by truncating the chain after just the first effective mode.
Following the treatment of Hughes \emph{et~al.} \cite{Hughes2009a}, this entails redefining our system Hamiltonian to include only the phonon operators $\hat Q_1$ and $\hat P_1$, i.e.
\begin{align}
    \label{eq:Hamiltonian sys with Q_1}
   \hat H_{\rm ep}^{(1)} &=\hbar\Omega_\mu(\hat{\sigma}^{+}_\mu+\hat{\sigma}^{-}_\mu)
   +\ket{X}\bra{X}(\hbar\Delta_\mu+\hbar D_1\hat{Q}_{1})\nonumber\\
   &+\frac{1}{2}[\hat{P}_{1}^2+\widetilde\omega_{1}^2\hat{Q}_{1}^2]\ ,
\end{align}
% UNITS
%  [Omega^2 Q^2] = J
% ->  [Q]  =  s J^(1/2)
%
%  [hbar D Q]  = J
%       [D Q]  = s^(-1)
% ->      [D]  = s^(-2)J^(-1/2)
%
%  [hbar D Q Q]  = J
%       [D Q Q] = s^(-1)
% ->       [D]  = s^(-3)J^(-1) ????
and relegating the remaining effective modes to a Markovian treatment.  The relegated modes constitute a residual phonon bath that induces phenomenological damping in the first effective mode.
Using  $J(\omega)=\sum_k|\lambda_k|^2\delta(\omega-\omega_k)$ with the spectral function for the longitudinal acoustic phonons $J(\omega)=\alpha_{p}\omega^{3}
e^{-\omega^2/2\omega_{b}^2}$, where $\alpha_p$ is a coupling parameter and $\omega_b$ is a high frequency cutoff \cite{Roy2011b}. The equations \eqs{eq:D definition} and \eqr{eq:tilde omega definition} are calculated using the following integrals,

\begin{align}
    D_1^{2} &=\int_{0}^{\infty} J(\omega)d\omega = 2\alpha_{p}\omega_{b}^{4} \\
    \widetilde\omega_1^{2}&=\int_{0}^{\infty} \omega^2 J(\omega)d\omega = 8\alpha_{p}\omega_{b}^{6} .
\end{align}

Following Ref.~\cite{Hughes2009a}, we use the technique known as Markovian closure, where the chain of effective phonon modes is terminated by its interaction with a Markovian system that represents the residual phonon bath, to obtain the master equation for the density operator describing the state, $\rho$, of the QD and the first effective mode as follows:
\begin{align}
    \label{eq: non-Markovian master equation}
    \frac{\partial\hat{\rho}}{\partial   t}
  &=\frac{1}{i\hbar}[\hat{H}_{\rm ep}^{(1)},\hat{\rho}]
  +\frac{\gamma_R}{2}\mathcal{L}(\hat\sigma^{-}_{\up})
  +\frac{\gamma_R}{2}\mathcal{L}(\hat\sigma^{-}_{\dn})\nonumber\\
     & +\frac{\gamma_{\rm ph}}{i\hbar}
      [\hat{Q}_1,[\hat{P}_1,\hat{\rho}(t)]_{+}]
      -\frac{2\gamma_{\rm ph}E_{\rm th}}{\hbar^2}[\hat{Q}_1, [\hat{Q}_1,\hat{\rho}(t)]],
\end{align}
where $\hat{H}_{\rm ep}^{(1)}$ is given by \eq{eq:Hamiltonian sys with Q_1}, $\mathcal{L}(\hat O)=2\hat O\hat\rho\hat O^\dagger-\hat O^\dagger\hat O\hat\rho-\hat\rho\hat O^\dagger\hat O$ is the Lindblad superoperator, $\gamma_{R}$ is the radiative decay rate from the exciton to each ground state, $\gamma_{\rm ph}$ is the friction coefficient that represents the coupling between the first effective mode and the residual phonon bath, $[\cdot,\cdot]$ and $[\cdot,\cdot]_+$ are the commutator and anticommutator, respectively, and
$E_{\rm th}$ is the mean equilibrium energy of the first effective mode given by \cite{Hughes2009a}
\bea
    E_{\rm th}=\frac{\hbar\widetilde\Omega_{1}}{2}\coth\left(\frac{\hbar\widetilde\Omega_{1}}{2k_{B}T}\right).
\eea

\subsection{Modelling the QD-nuclear spin interaction}

The third stage of the QHE, where the QD returns to its initial state
$\ket{1}$, is represented by panel (d) of
Fig.~\ref{fig:QD stages}. %(c) of Fig.~\ref{fig: simulation of stage 1}.
It involves a controlled electron-nuclear spin flip mediated by the spin polarised nuclei.
The coupling of an electronic spin to a nuclear spin reservoir is known as the central spin problem and is important for quantum memories and exploiting entanglement in quantum computing.
In our case, we are interested in the nuclear spin reservoir serving as an entropy sink for the electron in the QD.
The scheme is, in its ideal form, a unitary one which accumulates entropy in the nuclear spin reservoir in a way that can be compared with algorithmic cooling \cite{Baugh2005}.  The Hamiltonian describing the interaction between the QD electron and the nuclear spins in an external magnetic field is given by \cite{Taylor2003}
\begin{align}
        \label{eq:Ham n-e}
    \hat H_{\rm en} &=  g^*\mu_B B_0 \hat S_z +\sum_j\left[g_n\mu_n B_0\hat I_{z}^{(j)} + 2 a_j\hat S_z\hat I_{z}^{(j)}\right. \nonumber\\
    &+   \left. a_j (\hat S_+\hat I_{-}^{(j)}+\hat S_-\hat I_{+}^{(j)})\right]
\end{align}
where $B_0$ is the magnitude of a magnetic field directed along the $z$ axis, $j$ indexes each nucleus, $\hat S_\mu$ and $\hat I^{(j)}_\mu$ are the spin operators for the electron and $j$th nucleus where $\mu=z$, $+$ or $-$ labels a  component of spin, raising or lowering operator, respectively, $a_j=\frac{1}{2}A v_0|\psi(\mathbf{r}_j)|^2$ is the (real) one-electron hyperfine interaction coefficient due to a nucleus at position $\mathbf{r}_j$, $v_0$ is a volume of a unit cell, $\psi(\mathbf{r})$ is the wave function of the electron, and $A$ is the hyperfine contact interaction.
The magnitude of the effective magnetic field (in the $z$ direction) seen by the electron is given by
\[
    \hat B_{\rm eff}=B_0-\frac{1}{g^*\mu_B}\sum_j a_j\hat I_{z}^{(j)}
\]
and so
\begin{align*}
    \hat H_{\rm en} &= g^*\mu_B \hat B_{\rm eff} \hat S_z +g_n\mu_n B_0\sum_j\hat I_{z}^{(j)}\nonumber\\
     &+ \sum_j a_j(\hat I_{-}^{(j)}\hat S_++\hat I_{+}^{(j)}\hat S_-)\ .
\end{align*}
We set the value of $B_0$ to satisfy
\begin{align} \label{eq:stark shift}
     g^*\mu_B \ip{\hat B_{\rm eff}} = g_n\mu_n B_0\ ,
\end{align}
so that the eigenstates of $\hat S_z$ and $\hat I_z^{(j)}$  have approximately the same energy gap and the electron-nuclei interaction spin exchange is energy conserving.
Transforming to a rotating frame at angular frequency $g_n\mu_n B_0/\hbar$ and making the rotating wave approximation gives the effective Hamiltonian as
\begin{align} \label{eq:H_en^eff}
    \hat H_{\rm en}^{(\rm eff)} =
               \sum_j a_j (\hat I_{-}^{(j)}\hat S_++\hat I_{+}^{(j)}\hat S_-)\ .
\end{align}
We assume that $B_0$ satisfies the condition in \eq{eq:stark shift} throughout the operation of the heat engine.  Essentially the field $B_0$ counteracts the Overhauser shift in the electron states and results in degenerate electronic states as depicted in Fig.~\ref{fig:QD stages}.

\begin{figure}[b]
%\captionsetup{justification=RaggedRight}
   \includegraphics[width=0.50\textwidth]{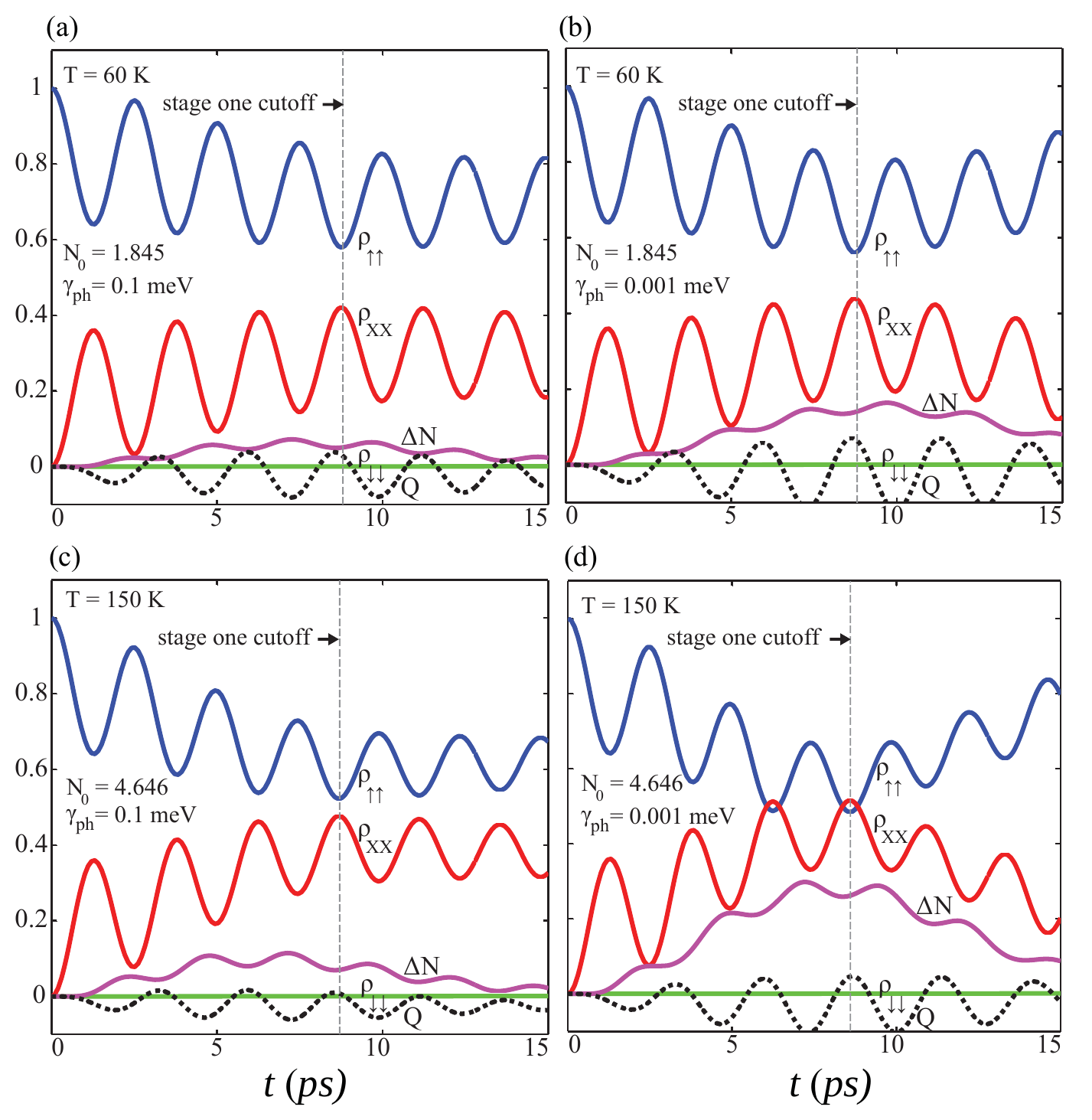}
  \caption{Simulations of the first stage (heat extraction) of the qdSHE.
  The panels show the dynamics for different combinations of temperature, $T=150$~K (upper panels) and $T=60$~K (lower panels), and friction coefficient, $\gamma_{\rm ph}=0.001$~meV (left panels) and $\gamma_{\rm ph}=0.1$~meV (right panels). We use $\hbar\Omega_{1}=0.75$ meV.
   The population of the ground state $\rho_{\up\up}$ (blue curves) reduces as the population of the exciton state $\rho_{XX}$ (red) increases.
   The change in the mean phonon number $\Delta \overline{N}_1$ (magenta) shows that heat is absorbed as the exciton population $\rho_{XX}$ increases.
  Oscillations in the mean position of the effective mode $\overline{Q}_1$ (black, dashed) are more damped for the higher friction coefficient (right panels).
    \label{fig: simulation of stage 1}}
\end{figure}

\section{Simulations of the \lowercase{qd}SHE}

We have simulated the operation of the qdSHE by numerically computing solutions of the master equation  \eq{eq: non-Markovian master equation} for a number of parameter regimes.
In all cases we use the following typical values \cite{Roy2011b} for self-assembled QD: $\alpha_{p}/(2\pi)^2=0.06~\mbox{ps}^2$, cutoff energy $\hbar\omega_{b}\approx1.48$~meV, and effective mode frequency $\widetilde\omega_{1}=\sqrt{5}\mbox{~meV}/\hbar$.
We examine each stage of the qdSHE separately below.

The solution $\hat{\rho}(t)$ of \eq{eq: non-Markovian master equation} allows us to monitor the evolution of the working fluid through the populations $\rho_{\up\up}$, $\rho_{\dn\dn}$ and $\rho_{XX}$ in the electron states $\ket{\up}$, $\ket{\dn}$ and $\ket{X}$, respectively, where
\begin{align}
        \rho_{\mu\mu}(t)=\tr(\ket{\mu}\bra{\mu}\hat{\rho}(t))
\end{align}
for $\mu=\up$, $\dn$ or $X$.  It also allows us to monitor the behavior of the effective mode.
We monitor the absorption of heat from the phonons by calculating the value of the mean phonon number, $\overline{N}(t)=\tr(\hat{N}\hat{\rho}(t))$, of the first effective mode over time.
The phonon number operator is given in terms of the mass-weighted coordinates $\hat{P}_1$ and $\hat{Q}_1$ by $\hat{N}=\hat{P}_1^2+\widetilde\omega_1^2\hat{Q}_1^2-\frac{1}{2}$.
The figures below display the change in the mean phonon number
\begin{align}
        \Delta\overline{N}(t)=\overline{N}(t)-\overline{N}_0
\end{align}
where $\overline{N}_0=\overline{N}(0)$. We also monitor the oscillations in the position of the effective mode by plotting
\begin{align}
        \overline{Q}_1(t)=\tr(\hat{Q}_1\hat{\rho}(t))\ .
\end{align}

\subsection{First stage: extracting heat}

The first stage entails irradiating the QD with a laser field of Rabi
frequency $\Omega_{1}=0.75\mbox{~meV}/\hbar$, and optical frequency
$\omega_{l}$ which is red detuned from the
$\ket{\uparrow}\leftrightarrow\ket{X}$ transition by $\Delta E/\hbar$,
where $\Delta E=2\mbox{~meV}$, as shown in
Fig.~\ref{fig:QD stages}(b).
The corresponding effective Rabi frequency is
$\Omega=\sqrt{\Delta^{2}+{\Omega_{1}}^{2}}\approx 2\mbox{~meV}/\hbar$
gives the highest frequency of the internal dynamics of the working
fluid. The associated period is commensurate with the phonon coherence
time, which is of order 1~ps, and this justifies our use of a
non-Markovian treatment of the phonons.

Figure~\ref{fig: simulation of stage 1} compares the ground state
population $\rho_{\up\up}$ (blue curve), exciton population
$\rho_{XX}$ (red curve), change in mean phonon number $\Delta
\overline{N}_1$ (magenta) and mean position of the effective mode
$\overline{Q}_1$ (black, dashed) for different combinations of
temperature $T$ and friction coefficient $\gamma_{\rm ph}$.  For
$T=60$~K, $\hbar\gamma_{\rm ph}=0.001$~meV, panel (a), almost $50\%$
percent of the ground state population is transferred to the exciton
state at $t\approx 10$~ps, and for $T=150$~K, panel (c), approximately
$60$\% population is transferred at a slightly earlier time.
Correspondingly, the change in mean phonon number $\Delta
\overline{N}_1$ reduces showing that heat is absorbed from the
phonons.  The non-Markovian nature of the phonons is manifested in
$\overline{Q}_1$ as oscillations about a drifting mean.  The absorbed
heat, population transfer and excursions in $\overline{Q}_1$ are all
less pronounced for the friction coefficient of $\gamma_{\rm
  ph}=0.1$~meV (right panels).

The optimum duration of the first stage corresponds to a pulse length
that gives high exciton population and phonon absorption. The first stage isn't stopped exactly at the maximum number of phonons absorbed as measurements of the exciton population are accessed easier experimentally. Thus we choose an exciton maxima that coincides closely to the maximum phonon absorption.

\subsection{Second stage: optical work output}

\begin{figure}[hb]
%\captionsetup{justification=RaggedRight}
   \includegraphics[width=0.50\textwidth]{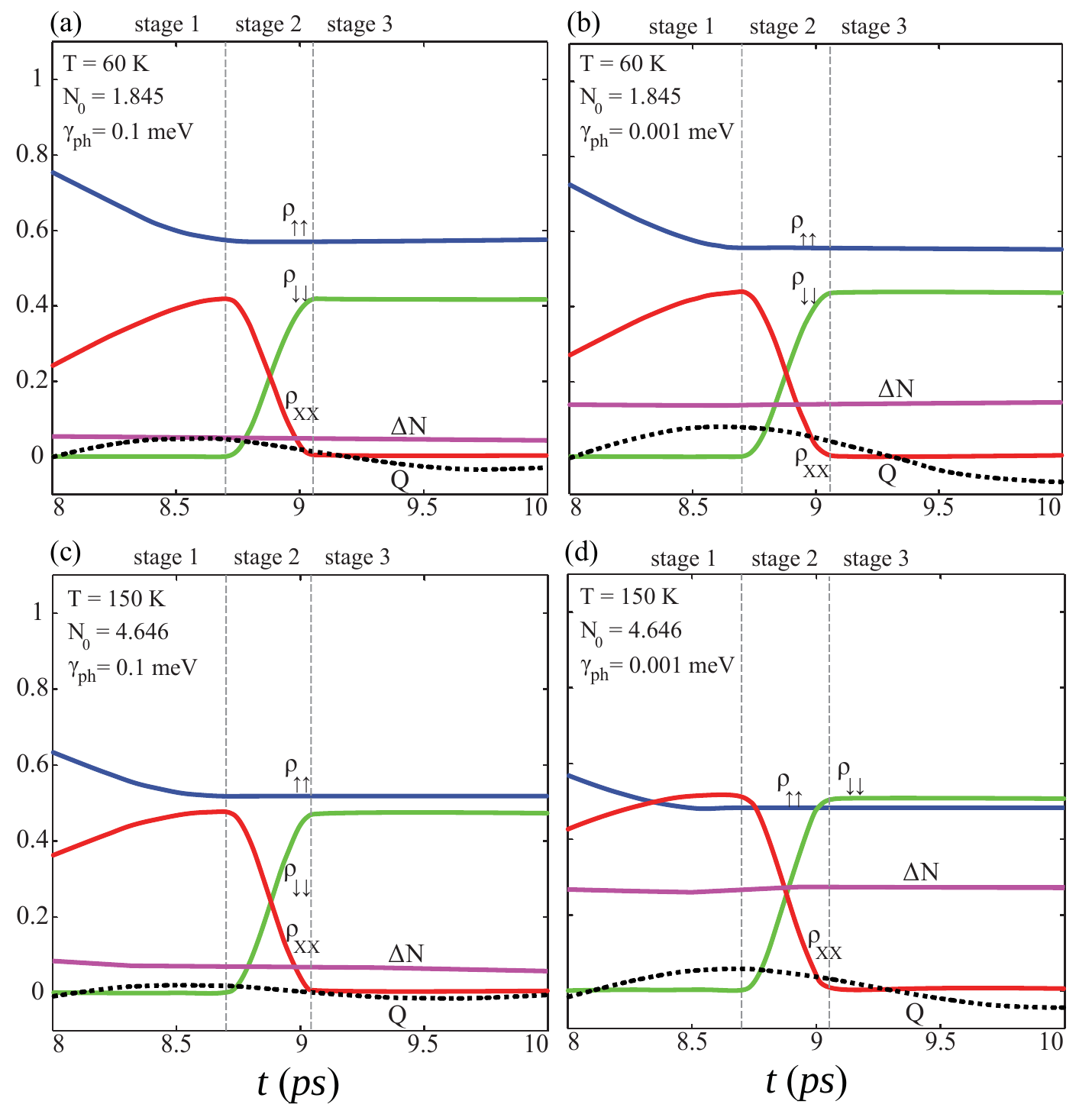}
    \caption{Simulation of stages 1 and 2.
     Stage 1 ends at $t\approx 8.7$~ps and stage 2 begins immediately for panels (a) and (b).
     The stage 1 ends $t\approx 9.1$~ps and stage 2 begins immediately for (c) and (d).
     Population transfer from $\rho_{XX}$ to $\rho_{\dn\dn}$. The green curve shows the population in state $\rho_{\dn\dn}$. Here we use
    $\hbar\Omega_{2}^{R}=4.316\mbox{~meV}$.
    \label{fig: simulation of stages 1 and 2}}
\end{figure}

The second stage entails a second laser field of Rabi frequency
$\Omega_{2}$ that is resonant with the zero phonon line of the
$\ket{\downarrow}\leftrightarrow \ket{X}$ transition, as illustrated
in Fig.~\ref{fig:QD stages}(c).
Figure \ref{fig: simulation of stages 1 and 2} shows a simulation of
stage two for the parameter
combination $T=60$~K and $\hbar\gamma_{\rm ph}=0.001$~meV.  To minimize
radiative decay from the exciton state, the second stage begins
immediately following the end of the first stage at the optimum time
of $t\approx 8.7$~ps.  The Rabi frequency $\Omega_{2}^{R}=4.316 \mbox{~meV}/\hbar$ is chosen to maximize the
population transfer from $\ket{X}$ to $\ket{\dn}$ and, as such, the
pulse area approximates a $\pi$ pulse.

The net result of the first two stages is that approximately $50\%$ of
the population is transferred from $\ket{\uparrow}$ to
$\ket{\downarrow}$. In this process, thermal phonon energy of $\Delta
E=2\mbox{~meV}$ is converted to coherent light by stimulated
emission. Loss due to the spontaneous emission is minimized since the
population of $\ket{X}$ is negligible at the end of the second stage.

\subsection{Third stage: information erasure}

We focus on the ideal case where the nuclear spins are initially fully polarised and, to demonstrate the underlying principle as simply as possible, we limit the analysis to nuclei with a spin quantum number of $I=1/2$.
We use the notation $\ket{n}_{\mathbf{x}}$ to represent a collective state of the nuclei in which $n$ nuclei are in the state $\ket\dn$ and the remainder are in $\ket\up$.
If $n>0$ the subscript $\mathbf{x}$, which is an $n$-dimensional vector, is used to uniquely specify one of many possible collective states with the same $z$ component of the total spin angular momentum.
We assume that each nucleus is initially polarised with a $z$ component of spin of $\hbar/2$ (i.e. in state $\ket\up$), and write the initial collective state of all the nuclei as $\ket{0}\equiv\ket{\up\up\up\cdots}$.

For the first cycle of the qdSHE, the time evolution of the electron-nuclear system follows \cite{Taylor2003}
\begin{align}
        \label{eq:state up,0 is a fixed point}
        \hat U^{\rm (eff)}_{\rm en}(t)\ket{\up,0} & = \ket{\up,0}\\
        \label{eq:state dn,0 evolution}
         \hat U^{\rm (eff)}_{\rm en}(t)\ket{\dn,0} & = \cos(\sqrt{\gamma} t) \ket{\dn,0} -i\sin(\sqrt{\gamma} t) \ket{\up,1}_0
\end{align}
where
\begin{align}
    \label{eq:hyperfine evolution}
        \hat U^{\rm (eff)}_{\rm en}(t) &=\exp[-i\hat H^{\rm (eff)}_{\rm en}t/\hbar]\\
    \label{eq:state 1}
        \ket{1}_0 &=\hat{\mathbb{I}}_-\ket{0}\\
    \label{eq:II-}
        \hat{\mathbb{I}}_- &\equiv \frac{1}{\sqrt{\gamma}}\sum_j a_j \hat{I}^{(j)}_-\\
    \label{eq:gamma}
        \gamma &\equiv {\sum}_j a_j^2
\end{align}
and $\ket{\updownarrow,n}_0$ represents the tensor product of electron state $\ket{\updownarrow}$ and nuclear state $\ket{n}_0$.
Notice that the state $\ket{\up,0}_0$ is a fixed point of the hyperfine interaction \eq{eq:H_en^eff}.
Choosing the duration $t=\pi/(2\sqrt{\gamma})$ results the electron spin being flipped to $\ket{\up}$ only if it is initially in the state $\ket{\dn}$ and remaining unchanged otherwise, which erases the memory of the electron, as represented by the mapping
\begin{subequations}
    \label{eq:1st partial erasure}
    \begin{align}
        \ket{\up,0} & \mapsto \ket{\up,0}\\
        \ket{\dn,0} & \mapsto-i\ket{\up,1}_0\ .
    \end{align}
\end{subequations}

The erasure stage is not complete, however, as the hyperfine interaction is ever present and the evolution described by \eq{eq:state dn,0 evolution} beyond the time $t=\pi/(2\sqrt{\gamma})$ would undo the erasure as the qdSHE undergoes its second cycle.
%This is true despite the fact that the first two stages of the second cycle can be completed well before the hyperfine interaction has time to affect the system.
To be more specific, at the end of stage two (panel (c) of Fig.~\ref{fig:QD stages}) of the second cycle the system would be in a statistical mixture of the states $\ket{\up,0}$, $\ket{\dn,0}$, $\ket{\up,1}_0$ and $\ket{\dn,1}_0$; in the case of the state $\ket{\up,1}_0$ the system would evolve beyond $t=\pi/(2\sqrt{\gamma})$ according to \eq{eq:state dn,0 evolution} back to $\ket{\dn,0}$, whereas the electron needs to remain unchanged in this case because it is in the desired state $\ket{\up}$.

To avoid this problem and complete the erasure stage at
$t=\pi/(2\sqrt{\gamma})$, we would like to perform an operation on the
nuclei that transforms both states $\ket{\up,0}$ and $\ket{\up,1}_0$
on the left sides of \eqs{eq:1st partial erasure} into \emph{fixed
  points} of the hyperfine interaction.  To minimize losses, the
desired operation should not incur a cost in terms of spinlabor or
work.

This leaves only the possibility of generating relative phase shifts
between the terms on the right side of \eq{eq:state 1}, which can be
implemented by briefly applying an additional magnetic field
$\mathbf{B}_{\rm pls}$ that is directed along the $z$ direction, i.e.
$\mathbf{B}_{\rm pls}(\mathbf{r}_j)=B_{\rm
  pls}(\mathbf{r}_j)\hat{\mathbf{k}}$ where $B_{\rm
  pls}(\mathbf{r}_j)$ is the magnitude of the field at the position
$\mathbf{r}_j$ of the $j$th nucleus.  If the duration of this magnetic
pulse is short compared to the spin flopping time of the order of
$t\sim 10$ ns, we can ignore the hyperfine interaction during the
pulse to a good approximation.  Experimentally magnetic pulses can be
generated as short as $3$ ps \cite{Wang2008} which easily satisfies
this condition.  We therefore model the evolution of the nuclei during
the pulse according to the Hamiltonian
\begin{align}
    \label{eq:Hamiltonian pulse}
        \hat H_{\rm pls}  &=  g_n\mu_n\sum_j B_{\rm pls}(\mathbf{r}_j)\hat I_{z}^{(j)}\ .
\end{align}
The effect of the pulse on the electron is the multiplication of its state by a trivial phase factor, which we ignore for brevity.
The effect on the nuclei is to transform the state $\ket{0}$ into $e^{-i\Theta\tau}\ket{0}$, and the state $\ket{1}_0$ into
\begin{align}
    \label{eq:state 1 tau}
       \ket{1}_\tau &= \hat U_{\rm pls}(\tau)\ket{1}_0
       =\frac{e^{-i\Theta\tau}}{\sqrt{\gamma}}\sum_j a_j e^{-i\theta_j\tau}I^{(j)}_-\ket{0}
\end{align}
where $\tau$ is the duration of the pulse, $\hat U_{\rm pls}(\tau)=e^{-i\hat H_{\rm pls}\tau/\hbar}$, $\theta_j\equiv g_n\mu_n B_{\rm pls}(\mathbf{r}_j)/\hbar$ and $\Theta\equiv\frac{1}{2}\sum_j\theta_j$.
The combination of the hyperfine interaction for $t=\pi/(2\sqrt{\gamma})$ followed by the magnetic pulse gives the mapping
\begin{subequations}
    \label{eq:1st erasure}
    \begin{align}
        \ket{\up,0} & \mapsto e^{-i\Theta\tau}\ket{\up,0}\\
        \ket{\dn,0} & \mapsto-i\ket{\up,1}_\tau\ .
    \end{align}
\end{subequations}
In Appendix \ref{appendix B} we show that further time evolution under the hyperfine interaction results in
\begin{align}
    &\hat U^{\rm (eff)}_{\rm en}(t)\ket{\up,1}_\tau
    = \ket{\up,1}_\tau +e^{-i\Theta\tau}\frac{\tilde\gamma(\tau)}{\gamma}\nonumber\\
        &\quad\times\left\{[\cos(\sqrt{\gamma} t)-1]\ket{\up,1}_0
             - i\sin(\sqrt{\gamma}t)\ket{\dn,0}\right\}
\end{align}
where
\begin{align}
    \label{eq:gamma tilde}
        \tilde{\gamma}(\tau) & \equiv\sum_j a_j^2 e^{-i\theta_j\tau}\ .
\end{align}
For suitable choices of the magnetic field $B_{\rm pls}(\mathbf{r}_j)$ and the pulse duration $\tau$ the factor $\tilde\gamma(\tau)/\gamma$ is negligible.
In that case, the state $\ket{\up,1}_\tau$ is an approximate fixed point of the hyperfine interaction.
As $\ket{\up,0}$  is also a fixed point, the erasure process is effectively halted by the magnetic pulse.

We now estimate the condition on the magnetic field and pulse
duration as follows. Using $a_j=\frac{1}{2}A
v_0|\psi(\mathbf{r}_j)|^2$, taking the wave function of the electron
to be a spherically symmetric Gaussian,
i.e. $\psi(x,y,z)=1/(2\pi\sigma^2)^{3/4}
\exp[-(x^2+y^2+z^2)/(4\sigma^2)]$ where $\sigma$ is the standard
deviation, treating the magnetic field to vary linearly along the $x$
direction only, i.e. $B_{\rm pls}(\mathbf{r}_j)=b_{\rm pls} x_j + C$
where $b_{\rm pls}$ and $C$ are constants and $x_j$ is the $x$
coordinate of $\mathbf{r}_j$, and approximating the sum over $j$ as a
volume integral yields, from \eq{eq:gamma tilde},
\begin{align}
    \tilde{\gamma}(\tau)
        &\approx \int \frac{A^2 v_0^2}{4(2\pi\sigma^2)^3}
        e^{-(x^2+y^2+z^2)/\sigma^2 -i \phi\tau x} dxdydz\\
        &= \frac{A^2 v_0^2}{32\pi^{3/2}\sigma^3} e^{-\phi^2\tau^2\sigma^2/4}
\end{align}
where $\phi \equiv g_n\mu_n b_{\rm pls}/\hbar$ and, for brevity, we have ignored the overall phase factor due to the constant $C$ of the magnetic field.
For comparison, the corresponding calculation of $\gamma$ using \eq{eq:gamma} is
\begin{align}
    \tilde{\gamma}(\tau)
        &\approx \int \frac{A^2 v_0^2}{4(2\pi\sigma^2)^3}
        e^{-(x^2+y^2+z^2)/\sigma^2} dxdydz\\
        &=\frac{A^2 v_0^2}{32\pi^{3/2}\sigma^3}
\end{align}
and so
\begin{align}
        \frac{\tilde{\gamma}(\tau)}{\gamma} &\approx e^{-\phi^2\tau^2\sigma^2/4}\ .
\end{align}
For the magnetic pulse to be effective, the right side should to be negligible compared to unity, and so we need $\phi\tau\gg 1/(2\sigma)$, and thus
\begin{align}
    \label{eq:condition for magnetic pulse}
        b_{\rm pls}\tau\gg \frac{\hbar}{2 g_n\mu_n\sigma}\ .
\end{align}

In principle, there is no fundamental reason that would prevent this condition from being achievable, but whether it can be satisfied in practice is a question of the available technology.
As an example, consider the magnetic field generated using pulsed electrical currents.
Imagine a current of $I$ is carried by a nanofabricated wire that has a circular cross-section of radius $R$.
Linearising the formula for the magnetic field surrounding a long straight wire gives $b_{\rm pls}=\mu_0 I/[2\pi (R+r)^2]$ as an estimate of the field gradient at a distance from $r$ from its surface.
Hence, in terms of the current pulse the condition \eq{eq:condition for magnetic pulse} becomes
\begin{align}
    \label{eq:condition for current pulse}
        I\tau\gg \frac{\hbar 2\pi (R+r)^2}{2 g_n\mu_0\mu_n\sigma}\ .
\end{align}
Using $R+r=2\sigma\approx 10$~nm and $g_n\approx 5$ (proton value) gives
\begin{align}
    \label{eq:condition for current pulse}
        I\tau\gg 4\times 10^{-10}\mbox{ As}\ .
\end{align}

To be consistent with our assumption that the duration of the pulse is much less than the spin flopping time of $10$~ns, we set $\tau\approx 1$~ns.
This gives a condition on the current of $I\gg 0.4$~A which appears feasible given that Wang \emph{et al.} \cite{Wang2008} have produced a current pulse of 20~A that lasted 3~ps.
However, their experiment used a conductor with a cross-section that is much larger than that assumed here.
The difficulty with a wire conductor with a radius of 10~nm is ohmic heating due to its high resistance.
This issue would have to be addressed before the erasure scheme was experimentally feasible.

The same erasure process, i.e. evolution described by the combination $\hat U_{\rm pls}(\tau)\hat U^{\rm (eff)}_{\rm en}(t)$ for $t=\pi/(2\sqrt{\gamma})$ and an appropriate value of $\tau$, is applied in the third stage of every cycle.
This leads to the possibility that $1\hbar$ of spin angular momentum is transferred from the nuclei to the electron, as represented by the lowering operator $\hat{\mathbb{I}}_-$ in  \eq{eq:state 1}, in each cycle.
Taking account of the evolution $\hat U_{\rm pls}(\tau)$ due to the magnetic pulse then suggests that after $m$ cycles, the nuclei will be in a mixture of states of the kind $\ket{n}_{\mathbf{t}}$ for $n\le m$, where
\begin{align}
    \label{eq:definition of state n}
     \ket{n}_{\mathbf{t}} &= \prod_{j=1}^{n}\left[\hat U_{\rm pls}(t_j)\hat{\mathbb{I}}_-\right]\ket{0}
\end{align}
and $\mathbf{t}=(t_1,t_2,\ldots,t_n)$ is an $n$-dimensional vector of various combinations of pulse durations.
The order of the factors in the product in \eq{eq:definition of state n} is defined to follow $\prod_{j=1}^{n}\hat Q_j=\hat Q_n \hat Q_{n-1}\ldots \hat Q_1$.
In Appendix \ref{appendix B} we show that, provided $n$ is relatively small compared to the total number of nuclei involved in the hyperfine interaction, the time evolution of $\ket{n}_{\mathbf{t}}$ is approximated by
\begin{align}
        \label{eq:state up,n is a fixed point}
    \hat U^{\rm (eff)}_{\rm en}(t)\ket{\up,n}_{\mathbf{t}} & = \ket{\up,n}_{\mathbf{t}} \\
        \label{eq:state dn,n evolution}
    \hat U^{\rm (eff)}_{\rm en}(t)\ket{\dn,n}_{\mathbf{t}} & = \cos(\sqrt{\gamma} t) \ket{\dn,n}_{\mathbf{t}} \nonumber\\
    &\qquad -i\sin(\sqrt{\gamma} t) \ket{\up,n+1}_{(\mathbf{t},0)} \ ,
\end{align}
to order $(\tilde{\gamma}(\tau)/\gamma)$ where
\begin{align}
    \label{eq:(t,0)}
        (\mathbf{t},0)\equiv(t_1,t_2,\ldots,t_n,0)
\end{align}
is an $(n+1)$-dimensional vector resulting from the concatenation of $\mathbf{t}$ with the 1-dimensional zero vector $(0)$.
\eqs{eq:state up,n is a fixed point} and \eqr{eq:state dn,n evolution} are in same form as \eqs{eq:state up,0 is a fixed point} and \eqr{eq:state dn,0 evolution}.  Applying the magnetic pulse at $t=\pi/(2\sqrt{\gamma})$ yields the approximate mapping to order $\tilde{\gamma}(\tau)/\gamma$:
\begin{subequations}
    \label{eq:n th erasure}
    \begin{align}
        \ket{\up,n}_{\mathbf{t}} & \mapsto e^{-i\Theta\tau}\ket{\up,n}_{\mathbf{t}'}\\
        \ket{\dn,n}_{\mathbf{t}} & \mapsto-i\ket{\up,n+1}_{(\mathbf{t},\tau)}\ .
    \end{align}
\end{subequations}
where $\mathbf{t}'=(t_1,t_2,\ldots,t_n+\tau)$.
The first and second stages of the $(m+1)$-th cycle transfer entropy from the phonon modes to the electron leaving it in a mixture of $\ket{\up}$ and $\ket{\dn}$.
This entropy is transferred to the nuclei in the third stage as described by \eqs{eq:n th erasure} with appropriate adjustments in the state of the nuclei.
Taken together, \eqs{eq:1st erasure} and \eqr{eq:n th erasure} confirm that the states $\ket{n}_{\mathbf{t}}$ defined in \eq{eq:definition of state n} do, in fact, represent the possible states of the nuclei generated by the operation of the qdSHE, provided the value of $m$ is small compared to the number of nuclei involved.

\section{Discussions and conclusions}

We can summarise the operating principle of the qdSHE rather clearly in terms of the conceptual diagram in Fig.~\ref{fig: comparison of Carnot engine with a SHE}(b).
The phonon bath of the crystal supporting the QD plays the role of the thermal reservoir and the nuclear spins play the role of the spin reservoir.
The relative amounts of heat $Q$, work $W$, spinlabor $\mathcal{L}_s$ and spintherm $\mathcal{Q}_s$ involved in the operation of the qdSHE are as follows.
For each photon of energy $\hbar\omega_l$ absorbed in the heat extraction stage Fig.~\ref{fig:QD stages}(b), a photon of energy $\hbar\omega_h$ is emitted in the optical work output stage Fig.~\ref{fig:QD stages}(c), giving an optical work $W$ output equal to the net increase in the total energy of the coherent laser fields, i.e. $W=\hbar(\omega_h-\omega_l)$.
By conservation of energy, the same quantity of heat $Q=\hbar(\omega_h-\omega_l)$ is absorbed from the thermal reservoir.
There are corresponding changes in the angular momentum as well (all amounts here are implicitly in terms of the $z$ component).
That is, at the end of the optical work extraction stage, the angular momentum of the electron in the QD is $1\hbar$ lower than it was at the beginning of the cycle Fig.~\ref{fig:QD stages}(a) and so, by conservation of angular momentum, the total angular momentum of the coherent laser fields is larger by $1\hbar$.
This increase represents $\mathcal{L}_S=-1\hbar$ of spinlabor being done on the QD by the laser fields.
In the erasure stage Fig.~\ref{fig:QD stages}(d), the nuclear spins lose $1\hbar$ of angular momentum in changing the state of the QD from $\ket{\dn}$ to $\ket{\up}$.
This loss represents the spin reservoir absorbing spintherm of $\mathcal{Q}_s=1\hbar$.
Hence we have $W=Q$ and $\mathcal{L}_s=-\mathcal{Q}_s$.
An important point is that changes in the coherent laser fields are interpreted as work $W$ and spinlabor $\mathcal{L}_s$, whereas changes in the reservoirs are interpreted as heat $Q$ and spintherm $\mathcal{Q}_s$.
Essentially, it is that fact that the laser fields are treated classically with well-defined amplitude and phase that allows them to be associated with work-like actions.
In contrast, the reservoirs are, in general, described by probabilistic properties, and so they are associated with heat-like quantities.

To conclude, it has only been relatively recently that attention has been given to the novel possibilities that arise from Jaynes' more generalised forms of statistical mechanics \cite{Jaynes1957,Jaynes1957b}.
The first to be studied was the cost of erasing information by Vaccaro and Barnett \cite{Vaccaro2011,Barnett2013}.  They showed that it need not be energy, as described by Landauer's erasure principle \cite{Landauer1961}, and that this could be used to design new types of heat engines.
In this paper we have presented a physical model of such a heat engine that operates between a thermal and a spin reservoir.
It draws heat from the thermal reservoir and uses a polarized spin reservoir as an
entropy sink without producing any waste heat.
In the process, spinlabor is dissipated as spintherm which is absorbed by the spin reservoir.
These results reinforce the importance of Jaynes' generalised framework and the potential it has for developing new ways of dealing with energy and other resources.

\section{Acknowledgements}
This research was supported by the ARC Linkage Grant No. LP140100797
and the Lockheed Martin Corporation.  We would like to acknowledge
helpful discussions with N. Allen, L. Uribarri, A. Jacombhood,
E. Streed, S.M Barnett and I. Adagideli.

\appendix
\section{Numerical implementation}
\label{appendix A}

\newcommand{\Hh}{\hat{H}}
\newcommand{\Qh}{\hat{Q}}
\newcommand{\rhoh}{\hat{\rho}}
\newcommand{\ibra}[1]{\langle#1|}
\newcommand{\iket}[1]{|#1\rangle}
\newcommand{\iout}[1]{\iket{#1}\ibra{#1}}
\newcommand{\iketbra}[2]{\iket{#1}\ibra{#2}}

For our analysis we need to solve equations (7) and (8)
or (10) and (8), depending on the stage. Equations (7) and (10)
can both be described in general form:
\begin{align}
  \Hh\equiv& (A+B\Qh_1)\iout{X} + \Hh_3 + \Hh_b.
  \label{eqn:HamGen}
\end{align}
Here $\Hh_3$ involves interactions only within the three-level
system and $\Hh_b$ involves interactions only within the bath.

Our numerical implementation involves truncating the Hilbert
space to avoid rapidly damped high-energy modes. We are thus
able to reproduce the important features of the model in a
fast code, with minimal numerical loss of accuracy.
The two main features of this scheme are discussed below.

\emph{Truncation of the bath}:
We first solve $\Hh_b$ analytically to find it solution as a set of
Harmonic oscillators which we denote $\iket{n}$ with energy
$E_n=S(n+\frac12)$. We then write
\begin{align}
  \Hh_b\approx& \sum_{0\leq n< N_c}E_n\iout{n},
  \\
  \Qh_1\approx& \sum_{0\leq n< N_c}\sum_{0\leq m< N_c}
  Q_{1,nm}\iketbra{n}{m}
\end{align}
where $N_c$ is a finite upper limit corresponding to trunctation
of energies greater than $E_c=S(N_c-\frac12)$.
Here $Q_{1,nm}=\ibra{n}\Qh_1\iket{m}$.

Equation \eqref{eqn:HamGen} can now be rewritten using the
product space $\iket{xn}\equiv\iket{x}\otimes\iket{n}$
involving all states in the three level system $\iket{x}\in
\{\iket{X},\iket{\uparrow},\iket{\downarrow}\}\equiv S_3$,
and the truncated space $\iket{n}$ for $0\leq n<N_c$
for the bath. Thus we may write
\begin{align}
  \Hh\approx &
  \sum_{x,y\in S_3}\sum_{0\leq n,m< N_c}
  H_{xy nm}\iket{xn}\ibra{ym}.
\end{align}
where, from the form of \eqref{eqn:HamGen} we see that
\begin{align}
  H_{xy nm}=&\delta_{xX}\delta_{yY}(A\delta_{nm} + BQ_{1,nm})
  \nonumber\\&
  + H_{3,xy}\delta_{nm} + \delta_{xy}\delta_{nm}E_n.
\end{align}
and $H_{3,xy}=\ibra{x}\Hh_3\iket{y}$.
We are thus able to reduce the full Hamiltonian $\Hh$ to a
$3N_c\times 3N_c$ matrix equation.

\emph{Semi-analytic evolution}:
Equation (8) involves evolution of the system's density matrix
$\rhoh$. Using the truncation scheme described above, we
write
\begin{align}
  \rhoh=&\sum_{x,y\in S_3}\sum_{0\leq n,m< N_c}
  \rho_{xynm}\iket{xn}\ibra{ym}
  \equiv \sum_I\rho_I\iket{xn}\ibra{ym}
\end{align}
where $\rho_{xynm}=\rho^*_{yxmn}$ follows from Hermiticity.
For the second expression we use
$I\equiv \iket{xn}\ibra{ym}$ for different combinations
of $x,y,n,m$ leading to $9N_c^2$ terms $\rho_I$.
After some work, the evolution equation (8) can be written as
\begin{align}
  \frac{\partial \rho_I}{\partial t}=&\sum_JV_{IJ}\rho_{J}.
  \label{eqn:EvV}
\end{align}

We now diagonalize $9N_c^2 \times 9N_c^2$ matrix
$V_{IJ}$ to find
\begin{align}
  \sum_JV_{IJ}\rho_{\kappa,J}=&v_{\kappa}\rho_{\kappa,I}
\end{align}
where $v_{\kappa}$ are the eigenvalues of $V_{IJ}$ and
$\rho_{\kappa,I}$ are the elements of the corresponding
eigenvectors. The properties of the evolution equation ensure
that $\Re[v_{\kappa}]<0$.
Now \eqref{eqn:EvV} can be solved directly giving
\begin{align}
  \rho_I(t)=&\sum_{\kappa} c_{\kappa}e^{v_{\kappa}(t-t_0)}\rho_{\kappa,I},
  &
  c_{\kappa}=&\sum_J \xi_{\kappa,J}\rho_J(t_0),
\end{align}
where $\xi_{\kappa,I}$ are the dual basis vectors for $\rho_{\kappa,I}$
obeying $\sum_I\xi_{\kappa,I}\rho_{\kappa',I}=\delta_{\kappa\kappa'}$,
and $\rho_I(t_0)=\ibra{xn}\rhoh(t_0)\iket{ym}$ are the
projected components of the density matrix at initial time $t_0$.

Clearly any eigenstate $\kappa$ with a large value for
$-\Re[v_{\kappa}]$ will contribute terms that decay quickly in
time from their initial value.
We may thus perform a second truncation here. Rather
than carrying out a full diagonalization of $V_{IJ}$
(requiring $O\boldsymbol{(}(9N_c^2)^3\boldsymbol{)}$ operations)
we seek only its eigenvalue/vector pairs with
$-\Re[v_{\kappa}]$ sufficiently small. There are typically
a relatively small number $N_t$ of these, allowing us to evaluate
this stage in $O\boldsymbol{(}N_t(9N_c^2)^2\boldsymbol{)}$ operations.

Combining both truncations allows us to solve the system
in a highly numerically tractable fashion, typically requiring
only a few minutes for a well-converged (against $N_c$)
run on a laptop computer. While the size of the active Hilbert
space considered here is fairly small by modern standards,
a more accurate understanding of the heat bath will require
several coupled head baths with a significantly larger
active space. Our efficient implementation
will allow improvements to the model to be studied in
these more complex systems.

\section{Details of the third stage}
\label{appendix B}

We first need to express the state of the nuclei, defined in \eq{eq:definition of state n}, more conveniently.
Expanding the product and judiciously factoring the unitary operators gives
\begin{align}
    \label{eq:expanding state n}
        \ket{n}_{\mathbf{t}}
        &= \left[\hat U_{\rm pls}(t_n)\hat{\mathbb{I}}_-\right]\!\!
           \left[\hat U_{\rm pls}(t_{n-1})\hat{\mathbb{I}}_-\right]\!\cdots\!
           \left[\hat U_{\rm pls}(t_{1})\hat{\mathbb{I}}_-\right]\!\!
           \ket{0}\nonumber\\
        &= \left[\hat U_{\rm pls}(T^{(\mathbf{t})}_n)\hat{\mathbb{I}}_-\hat U^\dagger_{\rm pls}(T^{(\mathbf{t})}_n)\right]\!\!\left[\hat U_{\rm pls}(T^{(\mathbf{t})}_{n-1})\hat{\mathbb{I}}_-\hat U^\dagger_{\rm pls}(T^{(\mathbf{t})}_{n-1})\right]\nonumber\\
           &\qquad\times\cdots\!\left[\hat U_{\rm pls}(T^{(\mathbf{t})}_{1})\hat{\mathbb{I}}_-\hat U^\dagger_{\rm pls}(T^{(\mathbf{t})}_{1})\right]\!\!\hat U_{\rm pls}(T^{(\mathbf{t})}_{1})
           \ket{0}\nonumber\\
        &= \prod_{m=1}^{n}\left[
        \hat U_{\rm pls}(T^{(\mathbf{t})}_m)\hat{\mathbb{I}}_-
             {\hat U}^\dagger_{\rm pls}(T^{(\mathbf{t})}_m)\right]
                      {\hat U}_{\rm pls}(T^{(\mathbf{t})}_1)
        \ket{0}
\end{align}
where
\begin{align}
        T^{(\mathbf{t})}_m\equiv
          \sum_{j=m}^{n}t_j
\end{align}
are partial sums of the elements of the $n$ dimensional vector $\mathbf{t}=(t_1,t_2,\ldots,t_n)$.
Recalling the definition of the collective lowering operator $\hat{\mathbb{I}}_-$ in \eq{eq:II-} and making use of $\hat U_{\rm pls}(\tau)=e^{-i\hat H_{\rm pls}\tau/\hbar}$, where $\hat H_{\rm pls}$ is given by \eq{eq:Hamiltonian pulse}, yields
\begin{align}
        \hat U_{\rm pls}(t)\hat{\mathbb{I}}_-
             {\hat U}^\dagger_{\rm pls}(t)
             &=\frac{1}{\sqrt{\gamma}}\sum_j a_j \hat U_{\rm pls}(t)\hat{I}^{(j)}_-{\hat U}^\dagger_{\rm pls}(t)\\
             &=\frac{1}{\sqrt{\gamma}}\sum_j a_j e^{i\theta_j t}  \hat{I}^{(j)}_-
\end{align}
where $\theta_j=g_n\mu_n B_{\rm pls}(r_j)\hbar$.
Inserting this into \eq{eq:expanding state n} and using the fact that ${\hat U}_{\rm pls}(t)\ket{0}=e^{-i\Theta t}\ket{0}$ where $\Theta=\frac{1}{2}\sum_j\theta_j$ then gives the desired form as
\begin{align}
            \label{eq:state n - product form}
        \ket{n}_{\mathbf{t}}
        &=    \frac{e^{-i\Theta T^{(\mathbf{t})}_1}}{\gamma^{n/2}}
            \prod_{m=1}^{n}\left[\sum_j a_j e^{i\theta_j T^{(\mathbf{t})}_m}  \hat{I}^{(j)}_-\right]\ket{0}\\
        &= \frac{e^{-i\Theta T^{(\mathbf{t})}_1}}{\gamma^{n/2}}
            \sum_{j,k,\ldots,\ell}\hat{\lbar}^{(j,n)}\hat{\lbar}^{(k,n-1)}
            \cdots\hat{\lbar}^{(\ell,1)}  \ket{0}
            \label{eq:state n - expanded as sum}
\end{align}
where the second line contains $n$ nested summations and, for convenience, we have defined
\begin{align}
    \label{eq:lbar}
        \hat{\lbar}^{(j,m)}\equiv a_j e^{i\theta_j T^{(\mathbf{t})}_m}\hat{I}^{(j)}_-\ .
\end{align}

Next, we need to evaluate the action of the hyperfine interaction $\hat U^{\rm (eff)}_{\rm en}(t)$ in \eq{eq:hyperfine evolution}, for which the  series expansion is
\begin{align}
        \hat U^{\rm (eff)}_{\rm en}(t)
        &= \sum_{m=0}^{\infty}\frac{1}{m!}\left[-i\frac{\sqrt{\gamma}t}{\hbar}
        \left(\hat{\mathbb{I}}_-\hat S_+ + \hat{\mathbb{I}}_+\hat S_-\right)\right]^m,
\end{align}
on the states $\ket{\up,n}_{\mathbf{t}}$ and $\ket{\dn,n}_{\mathbf{t}}$.
Taking account of the fact that, when acting on the state $\ket{\up}$, the only nonzero products of the operators $\hat{S}_+$ and $\hat{S}_-$ are given by $\ket{\dn}=\hat{S}_-\ket{\up}$, $\ket{\up}=\hat{S}_+\hat{S}_-\ket{\up}$, $\ket{\dn}=\hat{S}_-\hat{S}_+\hat{S}_-\ket{\up}$, $\ldots$, allows us to write
\begin{align}
    &\hat U^{\rm (eff)}_{\rm en}(t)\ket{\up,n}_{\mathbf{t}}\nonumber\\
        &\quad = \left[\sum_{m=0}^{\infty} \frac{(-1)^m}{(2m)!}\!\!\left(\frac{\sqrt{\gamma}t}{\hbar}\right)^{\!\!2m}
        \!\!\left(\hat{\mathbb{I}}_-\hat{\mathbb{I}}_+\right)^{\!\!m} \right]\!\ket{\up,n}_{\mathbf{t}}\nonumber\\
        &\qquad -i\frac{\sqrt{\gamma}t}{\hbar}\hat{\mathbb{I}}_+
        \!\!\left[\sum_{m=0}^{\infty} \frac{(-1)^m}{(2m+1)!}\!\!\left(\frac{\sqrt{\gamma}t}{\hbar}\right)^{\!\!2m}
        \!\!\left(\hat{\mathbb{I}}_-\hat{\mathbb{I}}_+\right)^{\!\!m} \right]\!\ket{\dn,n}_{\mathbf{t}}\ .
    \label{eq:evolution of up, n}
\end{align}
Similarly, an analogous consideration with respect to the state $\ket{\dn}$ leads to
\begin{align}
    &\hat U^{\rm (eff)}_{\rm en}(t)\ket{\dn,n}_{\mathbf{t}}\nonumber\\
        &\quad = \left[\sum_{m=0}^{\infty} \frac{(-1)^m}{(2m)!}\!\!\left(\frac{\sqrt{\gamma}t}{\hbar}\right)^{\!\!2m}
        \!\!\left(\hat{\mathbb{I}}_+\hat{\mathbb{I}}_-\right)^{\!\!m} \right]\!\ket{\dn,n}_{\mathbf{t}}\nonumber\\
        &\qquad -i\frac{\sqrt{\gamma}t}{\hbar}\hat{\mathbb{I}}_-
        \!\!\left[\sum_{m=0}^{\infty} \frac{(-1)^m}{(2m+1)!}\!\!\left(\frac{\sqrt{\gamma}t}{\hbar}\right)^{\!\!2m}
        \!\!\left(\hat{\mathbb{I}}_+\hat{\mathbb{I}}_-\right)^{\!\!m} \right]\!\ket{\up,n}_{\mathbf{t}}\ .
    \label{eq:evolution of dn, n}
\end{align}
These results reduce the problem to calculating expressions for terms such as $\hat{\mathbb{I}}_+\ket{n}_{\mathbf{t}}$, $(\hat{\mathbb{I}}_-\hat{\mathbb{I}}_+)^m\ket{n}_{\mathbf{t}}$, $\hat{\mathbb{I}}_-\ket{n}_{\mathbf{t}}$, and $(\hat{\mathbb{I}}_+\hat{\mathbb{I}}_-)^m\ket{n}_{\mathbf{t}}$.

We treat $\hat{\mathbb{I}}_-\ket{n}_{\mathbf{t}}$ first.
It follows from the definition of $\ket{n}_{\mathbf{t}}$ in \eq{eq:definition of state n} that
\begin{align}
    \label{eq:II- state n final}
    \hat{\mathbb{I}}_-\ket{n}_{\mathbf{t}}
         &= \hat U_{\rm pls}(0)\hat{\mathbb{I}}_-
            \prod_{j=1}^{n}\left[\hat U_{\rm pls}(t_j)\hat{\mathbb{I}}_-\right]\ket{0}\\
         &= \ket{n+1}_{(\mathbf{t},0)}
\end{align}
where $(\mathbf{t},0)$ is defined in \eq{eq:(t,0)} as the concatenation of $\mathbf{t}$ with  $(0)$.

The analysis of $\hat{\mathbb{I}}_+\ket{n}_{\mathbf{t}}$ is a little more involved.
For this note that as $\hat{I}^{(r)}_-\hat{I}^{(r)}_-=0$ for any value of $r$, the nonzero values of the product $\hat{I}^{(j)}_-\hat{I}^{(k)}_-\cdots\hat{I}^{(\ell)}_-$ of $n$ operators have the property that the values of the indices $j,k,\ldots,\ell$ form a unique set of positive integers and, in particular, this restricts $j$ from taking on $(n-1)$ values.
Focusing on these nonzero products only, we find
\begin{align}
     &\hat{I}^{(s)}_+\left(\hat{I}^{(j)}_-\hat{I}^{(k)}_-\cdots\hat{I}^{(\ell)}_-\right)\ket{0}\nonumber\\
     &=\left(\delta_{s,j}\!\!\left.\hat{\upsilon}^{(j)}\prod \hat{I}^{(\cdot)}_-\right|_{\hat{I}^{(j)}_-\mapsto\hat 1}
      +\delta_{s,k}\hat{\upsilon}^{(k)}\!\!\left.\prod \hat{I}^{(\cdot)}_-\right|_{\hat{I}^{(k)}_-\mapsto\hat 1} + \ldots    \right.\nonumber\\
      &\qquad\left.  +\delta_{s,\ell}\hat{\upsilon}^{(\ell)}\!\!\left.\prod \hat{I}^{(\cdot)}_-\right|_{\hat{I}^{(\ell)}_-\mapsto\hat 1}\right)\ket{0}
      \label{eq:I+I-I-...I-}
\end{align}
where the symbol $\prod \hat{I}^{(\cdot)}_-|_{\hat{I}^{(m)}_-\mapsto\hat 1}$ represents the product $\hat{I}^{(j)}_-\hat{I}^{(k)}_-\cdots\hat{I}^{(\ell)}_-$ with the factor $\hat{I}^{(m)}_-$ replaced with unity.
In deriving \eq{eq:I+I-I-...I-} we have used the facts that
both $\hat{I}^{(m)}_+$ and $\hat{I}^{(m)}_-$ commute with $\hat{I}^{(r)}_-$ for $m\ne r$,  $\hat{I}^{(m)}_+\hat{I}^{(r)}_-=(\ket{\up}\bra{\up})^{(r)}\equiv\hat{\upsilon}^{(r)}$ for $m=r$, and
 $\hat{I}^{(m)}_+\ket{0}=0$.
Notice that the right side of \eq{eq:I+I-I-...I-} retains the property of being zero unless the values of the indices $j,k,\ldots,\ell$ are unique.
Expanding $\hat{\mathbb{I}}_+\ket{n}_{\mathbf{t}}$ using the Hermitian conjugate of \eq{eq:II-} for $\hat{\mathbb{I}}_+$ and \eq{eq:state n - expanded as sum} for $\ket{n}_{\mathbf{t}}$, i.e.
\begin{align*}
        \hat{\mathbb{I}}_+\ket{n}_{\mathbf{t}}
        &=\left(\frac{1}{\sqrt{\gamma}}\sum_s a_s \hat{I}^{(s)}_+\right)\\
        &\quad\times\left(\frac{e^{-i\Theta T^{(\mathbf{t})}_1}}{\gamma^{n/2}}
            \sum_{j,k,\ldots,\ell}\hat{\lbar}^{(j,n)}\hat{\lbar}^{(k,n-1)}
            \cdots\hat{\lbar}^{(\ell,1)}  \ket{0}\right),
\end{align*}
recalling the definition $\hat{\lbar}^{(j,n)}\equiv a_j e^{i\theta_j T^{(\mathbf{t})}_m}\hat{I}^{(j)}_-$ in \eq{eq:lbar}, and then simplifying the result using \eq{eq:I+I-I-...I-} gives
\begin{align}
        \hat{\mathbb{I}}_+\ket{n}_{\mathbf{t}}
        &=\frac{e^{-i\Theta T^{(\mathbf{t})}_1}}{\gamma^{(n+1)/2}}
        \!\!\left(
            \sum_{j,k,\ldots,\ell} a^2_j e^{i\theta_j T^{(\mathbf{t})}_n}\hat{\upsilon}^{(j)}
            \!\!\left.\prod\hat{\lbar}^{(\cdot,\cdot)}_-\right|_{\hat{\lbar}^{(n,j)}_-\mapsto\hat 1}\right.\nonumber\\
        &\quad +    \sum_{j,k,\ldots,\ell} a^2_k e^{i\theta_k T^{(\mathbf{t})}_{n-1}}\hat{\upsilon}^{(k)}
            \!\!\left.\prod \hat{\lbar}^{(\cdot,\cdot)}_-\right|_{\hat{\lbar}^{(n-1,k)}_-\mapsto\hat 1}\nonumber\\
    \label{eq:II+ state n initial}
        &\quad \left.+\ldots
        + \sum_{j,k,\ldots,\ell} a^2_{\ell} e^{i\theta_{\ell} T^{(\mathbf{t})}_{1}}\hat{\upsilon}^{(\ell)}
            \!\!\left.\prod \hat{\lbar}^{(\cdot,\cdot)}_-\right|_{ \hat{\lbar}^{(1,\ell)}_-\mapsto\hat 1}\right)\!\!\ket{0}
\end{align}
where the symbol $\prod \hat{\lbar}^{(\cdot,\cdot)}_-|_{\hat{\lbar}^{(s,m)}_-\mapsto\hat 1}$ represents the product $\hat{\lbar}^{(n,j)}_-\hat{\lbar}^{(n-1,k)}_-\cdots\hat{\lbar}^{(1,\ell)}_-$ with the factor $\hat{\lbar}^{(s,m)}_-$ replaced with unity, which can be further simplified to
\begin{align}
    \label{eq:II+ state n intermediate}
    \hat{\mathbb{I}}_+\ket{n}_{\mathbf{t}}
        &=
        \sum_{m=1}^{n}
            \!\!\left[\!\frac{e^{-i\Theta t_1\delta_{m,1}}}{\gamma}
        \!\!\left(\!\!\sum_j a^2_j e^{i\theta_j T^{(\mathbf{t})}_m}\hat{\upsilon}^{(j)}\right)\!\!
            \ket{n\!-\!1}_{\mathbf{f}(\mathbf{t},m)}\right]
\end{align}
where $\mathbf{f}(\mathbf{t},m)=(f_1,f_2,\ldots,f_{n-1})$ is an $(n-1)$-dimensional vector derived from $\mathbf{t}$ according to
\begin{align}
    \label{eq:defn f(t,m)}
       f_j &=\left\{\begin{array}{l}
                  t_j \mbox{ for }  1\le j<m-1\ ,  \\
                  (t_{m-1}+t_{m}) \mbox{ for }  1\le j=m-1\ ,  \\
                  t_{j+1} \mbox{ for }  m \le j \le n-1\ .
                 \end{array}
        \right.
\end{align}
Further simplification is hindered by the fact the state $\ket{n-1}_{\mathbf{f}(\mathbf{t},m)}$ is not an eigenstate of $\hat{\upsilon}^{(j)}$.
However, \eq{eq:II+ state n intermediate} can be simplified \emph{approximately} provided $n$ is small compared to the total number of nuclear spins as follows.
In each summation in \eq{eq:II+ state n initial}, the state to the right of the operator $\hat{\upsilon}^{(s)}$ represents $(n-1)$ nuclei in the state $\ket{\dn}$ and the remainder in $\ket{\up}$, and so the result of the operator is zero for $(n-1)$ values of $s$ and unity otherwise.
The approximation entails replacing $\hat{\upsilon}^{(s)}$  with unity which leads to the final simplified result
\begin{align}
    \label{eq:II+ state n final}
        \hat{\mathbb{I}}_+\ket{n}_{\mathbf{t}}
        &=
        \sum_{m=1}^{n} e^{-i\Theta t_1\delta_{m,1}}\frac{\tilde{\gamma}(T^{(\mathbf{t})}_m)}{\gamma} \ket{n-1}_{\mathbf{f}(\mathbf{t},m)}\ ,
\end{align}
where $\tilde{\gamma}(t)$ is defined in \eq{eq:gamma tilde}, and incurs
a relative error of the order of $(n/N)$ where $N$ is the total number of nuclei involved in the hyperfine interaction.

It immediately follows from \eqs{eq:II- state n final} and \eqr{eq:II+ state n final} that
\begin{align}
    \label{eq:II-II+ state n}
      \hat{\mathbb{I}}_-\hat{\mathbb{I}}_+\ket{n}_{\mathbf{t}}
      &=
        \sum_{m=1}^{n} e^{-i\Theta t_1\delta_{m,1}}\frac{\tilde{\gamma}(T^{(\mathbf{t})}_m)}{\gamma} \ket{n}_{\mathbf{g}(\mathbf{t},m)}
\end{align}
where $\mathbf{g}(\mathbf{t},m)=(g_1,g_2,\ldots,g_n)$ is an $n$-dimensional vector given by $\mathbf{g}(\mathbf{t},m)=(\mathbf{f}(\mathbf{t},m),0)$, i.e.
\begin{align}
    \label{eq:defn g(t,m)}
       g_j &=\left\{\begin{array}{l}
                  t_j \mbox{ for }  1\le j<m-1\ ,  \\
                  (t_{m-1}+t_{m}) \mbox{ for }  1\le j=m-1\ ,  \\
                  t_{j+1} \mbox{ for }  m \le j \le n-1\ ,\\
                  0 \mbox{ for }   j =n\ .
                 \end{array}
        \right.
\end{align}
As discussed in the main text, we are interested in the regime where $\tilde{\gamma}(\tau)/\gamma$ is negligible.
As the values of $T^{(\mathbf{t})}_m$ are $\tau$ or larger, \eq{eq:II-II+ state n} shows that
\begin{align}
      \hat{\mathbb{I}}_-\hat{\mathbb{I}}_+\ket{n}_{\mathbf{t}}
      &=\mathcal{O}\left(\frac{\tilde{\gamma}(\tau)}{\gamma}\right)\ ,
\end{align}
and making use of \eqs{eq:II+ state n final} and \eqr{eq:II-II+ state n} multiple times then leads to
\begin{align}
    (\hat{\mathbb{I}}_-\hat{\mathbb{I}}_+)^m\ket{n}_{\mathbf{t}}
        &=\mathcal{O}\left(\frac{\tilde{\gamma}^m(\tau)}{\gamma^m}\right)\\
    \hat{\mathbb{I}}_+(\hat{\mathbb{I}}_-\hat{\mathbb{I}}_+)^m\ket{n}_{\mathbf{t}}
        &=\mathcal{O}\left(\frac{\tilde{\gamma}^{m+1}(\tau)}{\gamma^{m+1}}\right)\ ,
\end{align}
and substituting these results into \eq{eq:evolution of up, n} gives
\begin{align}
    \label{eq:U up n}
    &\hat U^{\rm (eff)}_{\rm en}(t)\ket{\up,n}_{\mathbf{t}}
    =\ket{\up,n}_{\mathbf{t}} +\mathcal{O}\left(\frac{\tilde{\gamma}(\tau)}{\gamma}\right)
\end{align}
which appears in the main text as \eq{eq:state up,n is a fixed point}.

It also follows from \eqs{eq:II- state n final} and \eqr{eq:II+ state n final} that
\begin{align}
    \hat{\mathbb{I}}_+\hat{\mathbb{I}}_-\ket{n}_{\mathbf{t}}
      &=\hat{\mathbb{I}}_+\ket{n+1}_{(\mathbf{t},0)}\\
      &=\sum_{m=1}^{n+1} e^{-i\Theta t_1\delta_{m,1}}
            \frac{\tilde{\gamma}(T^{(\mathbf{t}')}_m)}{\gamma} \ket{n}_{\mathbf{f}(\mathbf{t}',m)}\\
      &=\ket{n}_{\mathbf{t}}+\mathcal{O}\left(\frac{\tilde{\gamma}(\tau)}{\gamma}\right)
\end{align}
where $\mathbf{t}'=(\mathbf{t},0)$, and in the last line we have made use of the facts that $T^{(\mathbf{t}')}_{n+1}=0$ and $\mathbf{f}(\mathbf{t}',n+1)=\mathbf{t}$, thus
\begin{align}
    (\hat{\mathbb{I}}_+\hat{\mathbb{I}}_-)^m\ket{n}_{\mathbf{t}}
      &=\ket{n}_{\mathbf{t}}+\mathcal{O}\left(\frac{\tilde{\gamma}(\tau)}{\gamma}\right)\\
    \hat{\mathbb{I}}_-(\hat{\mathbb{I}}_+\hat{\mathbb{I}}_-)^m\ket{n}_{\mathbf{t}}
      &=\ket{n+1}_{(\mathbf{t},0)}+\mathcal{O}\left(\frac{\tilde{\gamma}(\tau)}{\gamma}\right)\ ,
\end{align}
and using these results in \eq{eq:evolution of dn, n} gives
\begin{align}
    \label{eq:U dn n}
    \hat U^{\rm (eff)}_{\rm en}(t)\ket{\dn,n}_{\mathbf{t}} & = \cos(\sqrt{\gamma} t) \ket{\dn,n}_{\mathbf{t}} \nonumber\\
    &\quad -i\sin(\sqrt{\gamma} t) \ket{\up,n+1}_{(\mathbf{t},0)} +\mathcal{O}\left(\frac{\tilde{\gamma}(\tau)}{\gamma}\right)\ .
\end{align}
This is the basis of \eq{eq:state dn,n evolution} in the main text.

Although the preceding analysis gives the details needed to support the calculations described in the text, it does not make clear how the magnetic pulse generates fixed points of the hyperfine interaction.
To address this issue we treat the simplest situation where the nuclei are in the state $\ket{\up,1}_0$.
This state occurs, potentially, during the first cycle before the magnetic pulse is applied.
Note that for any state to be a fixed point of the dynamics, it must be an eigenstate of the corresponding Hamiltonian.
We would, therefore, like to compare the action of the hyperfine
Hamiltonian $\hat H_{\rm en}^{(\rm eff)}$ on the state before and
after the magnetic pulse is applied; these states are given by
$\ket{\up,1}_0$ and $\hat{U}_{\rm pls}(\tau)\ket{\up,1}_0
=\ket{\up,1}_{\tau}$, respectively.

Using the expressions for $\hat H_{\rm en}^{(\rm eff)}$ in \eq{eq:H_en^eff}, $\ket{\up,1}_0$ in \eq{eq:state 1}, $\hat{\mathbb{I}}_{-}$ in \eq{eq:II-}, and $\hat{U}_{\rm pls}(\tau)\ket{\up,1}_0$ in \eq{eq:state 1 tau} we find
\begin{align}
       \hat H_{\rm en}^{(\rm eff)}\ket{\up,1}_0
           &=  \!\! \left(\sum_j a_j\hat{I}^{(j)}_+\right)\!\!\left(\frac{1}{\sqrt{\gamma}}\sum_k a_k\hat{I}^{(k)}_-\right)\ket{\dn,0}\nonumber\\
       \label{eq:compare state 1}
           &=\frac{1}{\sqrt{\gamma}}\left(\sum_k a_k^2\right)\ket{\dn,0}\ ,
\end{align}
and
\begin{align}
      \hat H_{\rm en}^{(\rm eff)}\hat{U}_{\rm pls}(\tau)\ket{\up,1}_0
           &=\!\!\left(\sum_j a_j\hat{I}^{(j)}_+\right)\nonumber\\
           &\quad \times\left(\frac{e^{-i\Theta\tau}}{\sqrt{\gamma}}\sum_k a_k e^{-i\theta_k\tau}I^{(k)}_-\right)\ket{\dn,0}\nonumber\\
       \label{eq:compare state 1 tau}
           &=\frac{e^{-i\Theta\tau}}{\sqrt{\gamma}}\left(\sum_k a_k^2 e^{-i\theta_k\tau}\right)\ket{\dn,0}\ .
\end{align}
As the parameters $a_k$ are real, the complex phase factors in the summand in the last line of \eq{eq:compare state 1 tau} allow the magnitude of the resulting sum to be much smaller than the corresponding sum in \eq{eq:compare state 1}.
The dynamics of the hyperfine interaction associated with \eq{eq:compare state 1 tau} would then be significantly suppressed in comparison to the dynamics associated with \eq{eq:compare state 1}.
This is the context in which the state $\hat{U}_{\rm pls}(\tau)\ket{\up,1}_0$ is approximately a fixed point of the hyperfine interaction.
In fact, the summations in \eqs{eq:compare state 1} and \eqr{eq:compare state 1 tau} are the parameters $\gamma$ and $\tilde{\gamma}(\tau)$ defined in \eqs{eq:gamma} and \eqr{eq:gamma tilde}, respectively, and the main text shows how $|\tilde{\gamma}(\tau)|\ll \gamma$ can be engineered.

\bibliography{heatengine}

\begin{thebibliography}{52}%
\makeatletter
\providecommand \@ifxundefined [1]{%
 \@ifx{#1\undefined}
}%
\providecommand \@ifnum [1]{%
 \ifnum #1\expandafter \@firstoftwo
 \else \expandafter \@secondoftwo
 \fi
}%
\providecommand \@ifx [1]{%
 \ifx #1\expandafter \@firstoftwo
 \else \expandafter \@secondoftwo
 \fi
}%
\providecommand \natexlab [1]{#1}%
\providecommand \enquote  [1]{``#1''}%
\providecommand \bibnamefont  [1]{#1}%
\providecommand \bibfnamefont [1]{#1}%
\providecommand \citenamefont [1]{#1}%
\providecommand \href@noop [0]{\@secondoftwo}%
\providecommand \href [0]{\begingroup \@sanitize@url \@href}%
\providecommand \@href[1]{\@@startlink{#1}\@@href}%
\providecommand \@@href[1]{\endgroup#1\@@endlink}%
\providecommand \@sanitize@url [0]{\catcode `\\12\catcode `\$12\catcode
  `\&12\catcode `\#12\catcode `\^12\catcode `\_12\catcode `\%12\relax}%
\providecommand \@@startlink[1]{}%
\providecommand \@@endlink[0]{}%
\providecommand \url  [0]{\begingroup\@sanitize@url \@url }%
\providecommand \@url [1]{\endgroup\@href {#1}{\urlprefix }}%
\providecommand \urlprefix  [0]{URL }%
\providecommand \Eprint [0]{\href }%
\providecommand \doibase [0]{http://dx.doi.org/}%
\providecommand \selectlanguage [0]{\@gobble}%
\providecommand \bibinfo  [0]{\@secondoftwo}%
\providecommand \bibfield  [0]{\@secondoftwo}%
\providecommand \translation [1]{[#1]}%
\providecommand \BibitemOpen [0]{}%
\providecommand \bibitemStop [0]{}%
\providecommand \bibitemNoStop [0]{.\EOS\space}%
\providecommand \EOS [0]{\spacefactor3000\relax}%
\providecommand \BibitemShut  [1]{\csname bibitem#1\endcsname}%
\let\auto@bib@innerbib\@empty
%</preamble>
\bibitem [{\citenamefont {Maxwell}(1902)}]{Maxwell}%
  \BibitemOpen
  \bibfield  {author} {\bibinfo {author} {\bibfnamefont {J.~C.}\ \bibnamefont
  {Maxwell}},\ }\href@noop {} {\emph {\bibinfo {title} {Theory of Heat}}}\
  (\bibinfo  {publisher} {Longmans, Green and Co.},\ \bibinfo {address}
  {London},\ \bibinfo {year} {1902})\BibitemShut {NoStop}%
\bibitem [{\citenamefont {Bennett}(1982)}]{Bennett1982}%
  \BibitemOpen
  \bibfield  {author} {\bibinfo {author} {\bibfnamefont {C.~H.}\ \bibnamefont
  {Bennett}},\ }\href {\doibase 10.1007/BF02084158} {\bibfield  {journal}
  {\bibinfo  {journal} {International Journal of Theoretical Physics}\ }\textbf
  {\bibinfo {volume} {21}},\ \bibinfo {pages} {905} (\bibinfo {year}
  {1982})}\BibitemShut {NoStop}%
\bibitem [{\citenamefont {Landauer}(1961)}]{Landauer1961}%
  \BibitemOpen
  \bibfield  {author} {\bibinfo {author} {\bibfnamefont {R.}~\bibnamefont
  {Landauer}},\ }\href@noop {} {\bibfield  {journal} {\bibinfo  {journal} {IBM
  J. Res. Develop}\ }\textbf {\bibinfo {volume} {5}},\ \bibinfo {pages} {183}
  (\bibinfo {year} {1961})}\BibitemShut {NoStop}%
\bibitem [{\citenamefont {Mandal}\ and\ \citenamefont
  {Jarzynski}(2012)}]{Mandal2012}%
  \BibitemOpen
  \bibfield  {author} {\bibinfo {author} {\bibfnamefont {D.}~\bibnamefont
  {Mandal}}\ and\ \bibinfo {author} {\bibfnamefont {C.}~\bibnamefont
  {Jarzynski}},\ }\href {\doibase 10.1073/pnas.1204263109} {\bibfield
  {journal} {\bibinfo  {journal} {Proceedings of the National Academy of
  Sciences}\ }\textbf {\bibinfo {volume} {109}},\ \bibinfo {pages} {11641}
  (\bibinfo {year} {2012})}\BibitemShut {NoStop}%
\bibitem [{\citenamefont {Strasberg}\ \emph {et~al.}(2017)\citenamefont
  {Strasberg}, \citenamefont {Schaller}, \citenamefont {Brandes},\ and\
  \citenamefont {Esposito}}]{Strasberg2017}%
  \BibitemOpen
  \bibfield  {author} {\bibinfo {author} {\bibfnamefont {P.}~\bibnamefont
  {Strasberg}}, \bibinfo {author} {\bibfnamefont {G.}~\bibnamefont {Schaller}},
  \bibinfo {author} {\bibfnamefont {T.}~\bibnamefont {Brandes}}, \ and\
  \bibinfo {author} {\bibfnamefont {M.}~\bibnamefont {Esposito}},\ }\href
  {\doibase http://dx.doi.org/10.1103/PhysRevX.7.021003} {\bibfield  {journal}
  {\bibinfo  {journal} {Phys. Rev. X}\ }\textbf {\bibinfo {volume} {7}},\
  \bibinfo {pages} {021003} (\bibinfo {year} {2017})}\BibitemShut {NoStop}%
\bibitem [{\citenamefont {Koski}\ \emph {et~al.}(2015)\citenamefont {Koski},
  \citenamefont {Kutvonen}, \citenamefont {Khaymovich}, \citenamefont
  {AlaNissila},\ and\ \citenamefont {Pekola}}]{Pekola2015}%
  \BibitemOpen
  \bibfield  {author} {\bibinfo {author} {\bibfnamefont {J.}~\bibnamefont
  {Koski}}, \bibinfo {author} {\bibfnamefont {A.}~\bibnamefont {Kutvonen}},
  \bibinfo {author} {\bibfnamefont {M.}~\bibnamefont {Khaymovich}}, \bibinfo
  {author} {\bibfnamefont {T.}~\bibnamefont {AlaNissila}}, \ and\ \bibinfo
  {author} {\bibfnamefont {J.}~\bibnamefont {Pekola}},\ }\href@noop {}
  {\bibfield  {journal} {\bibinfo  {journal} {Phys. Rev. Lett}\ }\textbf
  {\bibinfo {volume} {115}},\ \bibinfo {pages} {260602} (\bibinfo {year}
  {2015})}\BibitemShut {NoStop}%
\bibitem [{\citenamefont {Vidrighin}\ \emph {et~al.}(2016)\citenamefont
  {Vidrighin}, \citenamefont {Dahlsten}, \citenamefont {Barbieri},
  \citenamefont {Kim}, \citenamefont {Vedral},\ and\ \citenamefont
  {Walmsley}}]{Vidrighin2016}%
  \BibitemOpen
  \bibfield  {author} {\bibinfo {author} {\bibfnamefont {M.~D.}\ \bibnamefont
  {Vidrighin}}, \bibinfo {author} {\bibfnamefont {O.}~\bibnamefont {Dahlsten}},
  \bibinfo {author} {\bibfnamefont {M.}~\bibnamefont {Barbieri}}, \bibinfo
  {author} {\bibfnamefont {M.~S.}\ \bibnamefont {Kim}}, \bibinfo {author}
  {\bibfnamefont {V.}~\bibnamefont {Vedral}}, \ and\ \bibinfo {author}
  {\bibfnamefont {I.~A.}\ \bibnamefont {Walmsley}},\ }\href {\doibase
  10.1103/PhysRevLett.116.050401} {\bibfield  {journal} {\bibinfo  {journal}
  {Phys. Rev. Lett.}\ }\textbf {\bibinfo {volume} {116}},\ \bibinfo {pages}
  {050401} (\bibinfo {year} {2016})}\BibitemShut {NoStop}%
\bibitem [{\citenamefont {Jaynes}(1957{\natexlab{a}})}]{Jaynes1957}%
  \BibitemOpen
  \bibfield  {author} {\bibinfo {author} {\bibfnamefont {E.~T.}\ \bibnamefont
  {Jaynes}},\ }\href@noop {} {\bibfield  {journal} {\bibinfo  {journal} {Phys.
  Rev}\ }\textbf {\bibinfo {volume} {106}},\ \bibinfo {pages} {620} (\bibinfo
  {year} {1957}{\natexlab{a}})}\BibitemShut {NoStop}%
\bibitem [{\citenamefont {Jaynes}(1957{\natexlab{b}})}]{Jaynes1957b}%
  \BibitemOpen
  \bibfield  {author} {\bibinfo {author} {\bibfnamefont {E.~T.}\ \bibnamefont
  {Jaynes}},\ }\href@noop {} {\bibfield  {journal} {\bibinfo  {journal} {Phys.
  Rev}\ }\textbf {\bibinfo {volume} {108}},\ \bibinfo {pages} {171} (\bibinfo
  {year} {1957}{\natexlab{b}})}\BibitemShut {NoStop}%
\bibitem [{\citenamefont {Vaccaro}\ and\ \citenamefont
  {Barnett}(2006)}]{Vaccaro2006}%
  \BibitemOpen
  \bibfield  {author} {\bibinfo {author} {\bibfnamefont {J.~A.}\ \bibnamefont
  {Vaccaro}}\ and\ \bibinfo {author} {\bibfnamefont {S.~M.}\ \bibnamefont
  {Barnett}},\ }\href {http://aqis-conf.org/archives/aqis06/} {\bibfield
  {journal} {\bibinfo  {journal} {\emph{Generalization of Landauer's
  Information Erasure Principle}, Asian Conference on Quantum Information
  Science, 1-4 September, Beijing, China}\ } (\bibinfo {year} {2006})},\
  \bibinfo {note} {http://aqis-conf.org/archives/aqis06/}\BibitemShut {NoStop}%
\bibitem [{\citenamefont {Vaccaro}\ and\ \citenamefont
  {Barnett}(2011)}]{Vaccaro2011}%
  \BibitemOpen
  \bibfield  {author} {\bibinfo {author} {\bibfnamefont {J.~A.}\ \bibnamefont
  {Vaccaro}}\ and\ \bibinfo {author} {\bibfnamefont {S.~M.}\ \bibnamefont
  {Barnett}},\ }\href@noop {} {\bibfield  {journal} {\bibinfo  {journal} {Proc.
  R. Soc}\ }\textbf {\bibinfo {volume} {467}},\ \bibinfo {pages} {1770}
  (\bibinfo {year} {2011})}\BibitemShut {NoStop}%
\bibitem [{\citenamefont {Barnett}\ and\ \citenamefont
  {Vaccaro}(2013)}]{Barnett2013}%
  \BibitemOpen
  \bibfield  {author} {\bibinfo {author} {\bibfnamefont {S.~M.}\ \bibnamefont
  {Barnett}}\ and\ \bibinfo {author} {\bibfnamefont {J.~A.}\ \bibnamefont
  {Vaccaro}},\ }\href@noop {} {\bibfield  {journal} {\bibinfo  {journal}
  {Entropy}\ }\textbf {\bibinfo {volume} {15}},\ \bibinfo {pages} {4956}
  (\bibinfo {year} {2013})}\BibitemShut {NoStop}%
\bibitem [{\citenamefont {Croucher}\ \emph {et~al.}(2018)\citenamefont
  {Croucher}, \citenamefont {Wright}, \citenamefont {Carvalho}, \citenamefont
  {Barnett},\ and\ \citenamefont {Vaccaro}}]{Croucher2018}%
  \BibitemOpen
  \bibfield  {author} {\bibinfo {author} {\bibfnamefont {T.}~\bibnamefont
  {Croucher}}, \bibinfo {author} {\bibfnamefont {J.~S.~S.}\ \bibnamefont
  {Wright}}, \bibinfo {author} {\bibfnamefont {A.~R.~R.}\ \bibnamefont
  {Carvalho}}, \bibinfo {author} {\bibfnamefont {S.~M.}\ \bibnamefont
  {Barnett}}, \ and\ \bibinfo {author} {\bibfnamefont {J.~A.}\ \bibnamefont
  {Vaccaro}},\ }\href {https://arxiv.org/abs/1803.08619} {\bibfield  {journal}
  {\bibinfo  {journal} {\emph{Information erasure}, arXiv:1803.08619
  [quant-ph]}\ } (\bibinfo {year} {2018})},\ \bibinfo {note}
  {https://arxiv.org/abs/1803.08619}\BibitemShut {NoStop}%
\bibitem [{\citenamefont {Croucher}\ \emph {et~al.}(2017)\citenamefont
  {Croucher}, \citenamefont {Bedkihal},\ and\ \citenamefont
  {Vaccaro}}]{Toshio2017}%
  \BibitemOpen
  \bibfield  {author} {\bibinfo {author} {\bibfnamefont {T.}~\bibnamefont
  {Croucher}}, \bibinfo {author} {\bibfnamefont {S.}~\bibnamefont {Bedkihal}},
  \ and\ \bibinfo {author} {\bibfnamefont {J.~A.}\ \bibnamefont {Vaccaro}},\
  }\href@noop {} {\bibfield  {journal} {\bibinfo  {journal} {Phys. Rev. Lett}\
  }\textbf {\bibinfo {volume} {118}},\ \bibinfo {pages} {060602} (\bibinfo
  {year} {2017})}\BibitemShut {NoStop}%
\bibitem [{\citenamefont {Guryanova}\ \emph {et~al.}(2015)\citenamefont
  {Guryanova}, \citenamefont {Popescu}, \citenamefont {Short}, \citenamefont
  {Silva},\ and\ \citenamefont {Skrzypczyk}}]{PST11}%
  \BibitemOpen
  \bibfield  {author} {\bibinfo {author} {\bibfnamefont {Y.}~\bibnamefont
  {Guryanova}}, \bibinfo {author} {\bibfnamefont {S.}~\bibnamefont {Popescu}},
  \bibinfo {author} {\bibfnamefont {A.}~\bibnamefont {Short}}, \bibinfo
  {author} {\bibfnamefont {R.}~\bibnamefont {Silva}}, \ and\ \bibinfo {author}
  {\bibfnamefont {P.}~\bibnamefont {Skrzypczyk}},\ }\href {\doibase
  http://dx.doi.org/10.1038/ncomms12049} {\bibfield  {journal} {\bibinfo
  {journal} {Nat. Commun.}\ }\textbf {\bibinfo {volume} {7}},\ \bibinfo {pages}
  {12049} (\bibinfo {year} {2015})}\BibitemShut {NoStop}%
\bibitem [{\citenamefont {{Bozkurt}}\ \emph {et~al.}(2017)\citenamefont
  {{Bozkurt}}, \citenamefont {{Pekerten}},\ and\ \citenamefont
  {{Adagideli}}}]{Bozkurt2017}%
  \BibitemOpen
  \bibfield  {author} {\bibinfo {author} {\bibfnamefont {A.}~\bibnamefont
  {{Bozkurt}}}, \bibinfo {author} {\bibfnamefont {B.}~\bibnamefont
  {{Pekerten}}}, \ and\ \bibinfo {author} {\bibfnamefont {I.}~\bibnamefont
  {{Adagideli}}},\ }\href@noop {} {\bibfield  {journal} {\bibinfo  {journal}
  {arXiv e-print}\ } (\bibinfo {year} {2017})},\ \Eprint
  {http://arxiv.org/abs/1705.04985v1} {arXiv:1705.04985v1 [cond-mat.mes-hall]}
  \BibitemShut {NoStop}%
\bibitem [{\citenamefont {Yunger~Halpern}\ \emph {et~al.}(2016)\citenamefont
  {Yunger~Halpern}, \citenamefont {Faist}, \citenamefont {Oppenheim},\ and\
  \citenamefont {Winter}}]{Halpern2015}%
  \BibitemOpen
  \bibfield  {author} {\bibinfo {author} {\bibfnamefont {N.}~\bibnamefont
  {Yunger~Halpern}}, \bibinfo {author} {\bibfnamefont {P.}~\bibnamefont
  {Faist}}, \bibinfo {author} {\bibfnamefont {J.}~\bibnamefont {Oppenheim}}, \
  and\ \bibinfo {author} {\bibfnamefont {A.}~\bibnamefont {Winter}},\ }\href
  {\doibase http://dx.doi.org/10.1038/ncomms12051} {\bibfield  {journal}
  {\bibinfo  {journal} {Nat. Commun.}\ }\textbf {\bibinfo {volume} {7}},\
  \bibinfo {pages} {12051} (\bibinfo {year} {2016})}\BibitemShut {NoStop}%
\bibitem [{\citenamefont {Lostaglio}\ \emph {et~al.}(2017)\citenamefont
  {Lostaglio}, \citenamefont {Jennings},\ and\ \citenamefont
  {Rudolph}}]{Lostaglio2017}%
  \BibitemOpen
  \bibfield  {author} {\bibinfo {author} {\bibfnamefont {M.}~\bibnamefont
  {Lostaglio}}, \bibinfo {author} {\bibfnamefont {D.}~\bibnamefont {Jennings}},
  \ and\ \bibinfo {author} {\bibfnamefont {T.}~\bibnamefont {Rudolph}},\ }\href
  {http://stacks.iop.org/1367-2630/19/i=4/a=043008} {\bibfield  {journal}
  {\bibinfo  {journal} {New Journal of Physics}\ }\textbf {\bibinfo {volume}
  {19}},\ \bibinfo {pages} {043008} (\bibinfo {year} {2017})}\BibitemShut
  {NoStop}%
\bibitem [{\citenamefont {Mahan}(2000)}]{Mahan2000}%
  \BibitemOpen
  \bibfield  {author} {\bibinfo {author} {\bibfnamefont {G.}~\bibnamefont
  {Mahan}},\ }\href@noop {} {\emph {\bibinfo {title} {Many Particle Physics}}}\
  (\bibinfo {year} {2000})\ \bibinfo {note} {3ed (Plenum, New York,
  2000)}\BibitemShut {NoStop}%
\bibitem [{\citenamefont {Imamoglu}\ \emph {et~al.}(2003)\citenamefont
  {Imamoglu}, \citenamefont {Knill}, \citenamefont {Tian},\ and\ \citenamefont
  {Zolle}}]{Imamoglu2003}%
  \BibitemOpen
  \bibfield  {author} {\bibinfo {author} {\bibfnamefont {A.}~\bibnamefont
  {Imamoglu}}, \bibinfo {author} {\bibfnamefont {E.}~\bibnamefont {Knill}},
  \bibinfo {author} {\bibfnamefont {L.}~\bibnamefont {Tian}}, \ and\ \bibinfo
  {author} {\bibfnamefont {P.}~\bibnamefont {Zolle}},\ }\href@noop {}
  {\bibfield  {journal} {\bibinfo  {journal} {Phys. Rev. Lett}\ }\textbf
  {\bibinfo {volume} {91}},\ \bibinfo {pages} {017402} (\bibinfo {year}
  {2003})}\BibitemShut {NoStop}%
\bibitem [{\citenamefont {Lampel}(1968)}]{Lampel1968}%
  \BibitemOpen
  \bibfield  {author} {\bibinfo {author} {\bibfnamefont {G.}~\bibnamefont
  {Lampel}},\ }\href {\doibase http://dx.doi.org/10.1103/PhysRevLett.20.491}
  {\bibfield  {journal} {\bibinfo  {journal} {Phys. Rev. Lett.}\ }\textbf
  {\bibinfo {volume} {20}},\ \bibinfo {pages} {491} (\bibinfo {year}
  {1968})}\BibitemShut {NoStop}%
\bibitem [{\citenamefont {Christ}\ \emph {et~al.}(2007)\citenamefont {Christ},
  \citenamefont {Cirac},\ and\ \citenamefont {Giedke}}]{Christ2007}%
  \BibitemOpen
  \bibfield  {author} {\bibinfo {author} {\bibfnamefont {H.}~\bibnamefont
  {Christ}}, \bibinfo {author} {\bibfnamefont {J.~I.}\ \bibnamefont {Cirac}}, \
  and\ \bibinfo {author} {\bibfnamefont {G.}~\bibnamefont {Giedke}},\ }\href
  {\doibase http://dx.doi.org/10.1103/PhysRevB.75.155324} {\bibfield  {journal}
  {\bibinfo  {journal} {Phys. Rev. B}\ }\textbf {\bibinfo {volume} {75}},\
  \bibinfo {pages} {155324} (\bibinfo {year} {2007})}\BibitemShut {NoStop}%
\bibitem [{\citenamefont {Chekhovich}\ \emph {et~al.}(2010)\citenamefont
  {Chekhovich}, \citenamefont {Makhonin}, \citenamefont {Kavokin},
  \citenamefont {Krysa}, \citenamefont {Skolnick},\ and\ \citenamefont
  {Tartakovskii}}]{Chekhovich2010}%
  \BibitemOpen
  \bibfield  {author} {\bibinfo {author} {\bibfnamefont {E.~A.}\ \bibnamefont
  {Chekhovich}}, \bibinfo {author} {\bibfnamefont {M.~N.}\ \bibnamefont
  {Makhonin}}, \bibinfo {author} {\bibfnamefont {K.~V.}\ \bibnamefont
  {Kavokin}}, \bibinfo {author} {\bibfnamefont {A.~B.}\ \bibnamefont {Krysa}},
  \bibinfo {author} {\bibfnamefont {M.~S.}\ \bibnamefont {Skolnick}}, \ and\
  \bibinfo {author} {\bibfnamefont {A.~I.}\ \bibnamefont {Tartakovskii}},\
  }\href {\doibase http://dx.doi.org/10.1103/PhysRevLett.104.066804} {\bibfield
   {journal} {\bibinfo  {journal} {Phys. Rev. Lett.}\ }\textbf {\bibinfo
  {volume} {104}},\ \bibinfo {pages} {066804} (\bibinfo {year}
  {2010})}\BibitemShut {NoStop}%
\bibitem [{\citenamefont {Besombes}\ \emph {et~al.}(2001)\citenamefont
  {Besombes}, \citenamefont {Kheng}, \citenamefont {Marsal},\ and\
  \citenamefont {Mariette}}]{Besombes2001}%
  \BibitemOpen
  \bibfield  {author} {\bibinfo {author} {\bibfnamefont {L.}~\bibnamefont
  {Besombes}}, \bibinfo {author} {\bibfnamefont {K.}~\bibnamefont {Kheng}},
  \bibinfo {author} {\bibfnamefont {L.}~\bibnamefont {Marsal}}, \ and\ \bibinfo
  {author} {\bibfnamefont {H.}~\bibnamefont {Mariette}},\ }\href {\doibase
  http://dx.doi.org/10.1103/PhysRevB.63.155307} {\bibfield  {journal} {\bibinfo
   {journal} {Phys. Rev. B}\ }\textbf {\bibinfo {volume} {63}},\ \bibinfo
  {pages} {155307} (\bibinfo {year} {2001})}\BibitemShut {NoStop}%
\bibitem [{\citenamefont {Wilson-Rae}\ and\ \citenamefont
  {Imamoglu}(2002)}]{Wilson-Rae2002}%
  \BibitemOpen
  \bibfield  {author} {\bibinfo {author} {\bibfnamefont {I.}~\bibnamefont
  {Wilson-Rae}}\ and\ \bibinfo {author} {\bibfnamefont {A.}~\bibnamefont
  {Imamoglu}},\ }\href {\doibase http://dx.doi.org/10.1103/PhysRevB.65.235311}
  {\bibfield  {journal} {\bibinfo  {journal} {Phys. Rev. B}\ }\textbf {\bibinfo
  {volume} {65}},\ \bibinfo {pages} {235311} (\bibinfo {year}
  {2002})}\BibitemShut {NoStop}%
\bibitem [{\citenamefont {Roy}\ and\ \citenamefont
  {Hughes}(2011{\natexlab{a}})}]{Roy2011a}%
  \BibitemOpen
  \bibfield  {author} {\bibinfo {author} {\bibfnamefont {C.}~\bibnamefont
  {Roy}}\ and\ \bibinfo {author} {\bibfnamefont {S.}~\bibnamefont {Hughes}},\
  }\href@noop {} {\bibfield  {journal} {\bibinfo  {journal} {Phys. Rev. X}\
  }\textbf {\bibinfo {volume} {1}},\ \bibinfo {pages} {021009} (\bibinfo {year}
  {2011}{\natexlab{a}})}\BibitemShut {NoStop}%
\bibitem [{\citenamefont {McCutcheon}\ and\ \citenamefont
  {Nazir}(2013)}]{McCutcheon2013}%
  \BibitemOpen
  \bibfield  {author} {\bibinfo {author} {\bibfnamefont {D.}~\bibnamefont
  {McCutcheon}}\ and\ \bibinfo {author} {\bibfnamefont {A.}~\bibnamefont
  {Nazir}},\ }\href {\doibase http://dx.doi.org/10.1103/PhysRevLett.110.217401}
  {\bibfield  {journal} {\bibinfo  {journal} {Phys. Rev. Lett.}\ }\textbf
  {\bibinfo {volume} {110}},\ \bibinfo {pages} {217401} (\bibinfo {year}
  {2013})}\BibitemShut {NoStop}%
\bibitem [{\citenamefont {Roy}\ and\ \citenamefont
  {Hughes}(2011{\natexlab{b}})}]{Roy2011b}%
  \BibitemOpen
  \bibfield  {author} {\bibinfo {author} {\bibfnamefont {C.}~\bibnamefont
  {Roy}}\ and\ \bibinfo {author} {\bibfnamefont {S.}~\bibnamefont {Hughes}},\
  }\href {\doibase http://dx.doi.org/10.1103/PhysRevLett.106.247403} {\bibfield
   {journal} {\bibinfo  {journal} {Phys. Rev. Lett.}\ }\textbf {\bibinfo
  {volume} {106}},\ \bibinfo {pages} {247403} (\bibinfo {year}
  {2011}{\natexlab{b}})}\BibitemShut {NoStop}%
\bibitem [{\citenamefont {Timm}(2011)}]{Timm2011}%
  \BibitemOpen
  \bibfield  {author} {\bibinfo {author} {\bibfnamefont {C.}~\bibnamefont
  {Timm}},\ }\href {\doibase http://dx.doi.org/10.1103/PhysRevB.83.115416}
  {\bibfield  {journal} {\bibinfo  {journal} {Phys. Rev. B}\ }\textbf {\bibinfo
  {volume} {83}},\ \bibinfo {pages} {115416} (\bibinfo {year}
  {2011})}\BibitemShut {NoStop}%
\bibitem [{\citenamefont {Stock}\ \emph {et~al.}(2011)\citenamefont {Stock},
  \citenamefont {Dachner}, \citenamefont {Warming}, \citenamefont {Schliwa},
  \citenamefont {Lochmann}, \citenamefont {Hoffmann}, \citenamefont {Toropov},
  \citenamefont {Bakarov}, \citenamefont {Derebezov}, \citenamefont {Richter},
  \citenamefont {Haisler}, \citenamefont {Knorr},\ and\ \citenamefont
  {Bimberg}}]{Stock2011}%
  \BibitemOpen
  \bibfield  {author} {\bibinfo {author} {\bibfnamefont {E.}~\bibnamefont
  {Stock}}, \bibinfo {author} {\bibfnamefont {M.-R.}\ \bibnamefont {Dachner}},
  \bibinfo {author} {\bibfnamefont {T.}~\bibnamefont {Warming}}, \bibinfo
  {author} {\bibfnamefont {A.}~\bibnamefont {Schliwa}}, \bibinfo {author}
  {\bibfnamefont {A.}~\bibnamefont {Lochmann}}, \bibinfo {author}
  {\bibfnamefont {A.}~\bibnamefont {Hoffmann}}, \bibinfo {author}
  {\bibfnamefont {A.}~\bibnamefont {Toropov}}, \bibinfo {author} {\bibfnamefont
  {A.}~\bibnamefont {Bakarov}}, \bibinfo {author} {\bibfnamefont
  {I.}~\bibnamefont {Derebezov}}, \bibinfo {author} {\bibfnamefont
  {M.}~\bibnamefont {Richter}}, \bibinfo {author} {\bibfnamefont
  {V.}~\bibnamefont {Haisler}}, \bibinfo {author} {\bibfnamefont
  {A.}~\bibnamefont {Knorr}}, \ and\ \bibinfo {author} {\bibfnamefont
  {D.}~\bibnamefont {Bimberg}},\ }\href {\doibase
  http://dx.doi.org/10.1103/PhysRevB.83.041304} {\bibfield  {journal} {\bibinfo
   {journal} {Phys. Rev. B}\ }\textbf {\bibinfo {volume} {83}},\ \bibinfo
  {pages} {041304(R)} (\bibinfo {year} {2011})}\BibitemShut {NoStop}%
\bibitem [{\citenamefont {Majumdar}\ \emph {et~al.}(2012)\citenamefont
  {Majumdar}, \citenamefont {Bajcsy}, \citenamefont {Rundquist}, \citenamefont
  {Kim},\ and\ \citenamefont {Vuckovic}}]{Majumdar2012}%
  \BibitemOpen
  \bibfield  {author} {\bibinfo {author} {\bibfnamefont {A.}~\bibnamefont
  {Majumdar}}, \bibinfo {author} {\bibfnamefont {M.}~\bibnamefont {Bajcsy}},
  \bibinfo {author} {\bibfnamefont {A.}~\bibnamefont {Rundquist}}, \bibinfo
  {author} {\bibfnamefont {E.}~\bibnamefont {Kim}}, \ and\ \bibinfo {author}
  {\bibfnamefont {J.}~\bibnamefont {Vuckovic}},\ }\href {\doibase
  http://dx.doi.org/10.1103/PhysRevB.85.195301} {\bibfield  {journal} {\bibinfo
   {journal} {Phys. Rev. B}\ }\textbf {\bibinfo {volume} {85}},\ \bibinfo
  {pages} {195301} (\bibinfo {year} {2012})}\BibitemShut {NoStop}%
\bibitem [{\citenamefont {Kaer}\ \emph {et~al.}(2012)\citenamefont {Kaer},
  \citenamefont {Nielsen}, \citenamefont {Lodahl}, \citenamefont {Jauho},\ and\
  \citenamefont {Mork}}]{Kaer2012}%
  \BibitemOpen
  \bibfield  {author} {\bibinfo {author} {\bibfnamefont {P.}~\bibnamefont
  {Kaer}}, \bibinfo {author} {\bibfnamefont {T.~R.}\ \bibnamefont {Nielsen}},
  \bibinfo {author} {\bibfnamefont {P.}~\bibnamefont {Lodahl}}, \bibinfo
  {author} {\bibfnamefont {A.}~\bibnamefont {Jauho}}, \ and\ \bibinfo {author}
  {\bibfnamefont {J.}~\bibnamefont {Mork}},\ }\href {\doibase
  http://dx.doi.org/10.1103/PhysRevB.86.085302} {\bibfield  {journal} {\bibinfo
   {journal} {Phys. Rev. B}\ }\textbf {\bibinfo {volume} {86}},\ \bibinfo
  {pages} {085302} (\bibinfo {year} {2012})}\BibitemShut {NoStop}%
\bibitem [{\citenamefont {Weiler}\ \emph {et~al.}(2012)\citenamefont {Weiler},
  \citenamefont {Ulhaq}, \citenamefont {Ulrich}, \citenamefont {Richter},
  \citenamefont {Jetter}, \citenamefont {Michler}, \citenamefont {Roy},\ and\
  \citenamefont {Hughes}}]{Weiler2012}%
  \BibitemOpen
  \bibfield  {author} {\bibinfo {author} {\bibfnamefont {S.}~\bibnamefont
  {Weiler}}, \bibinfo {author} {\bibfnamefont {A.}~\bibnamefont {Ulhaq}},
  \bibinfo {author} {\bibfnamefont {S.~M.}\ \bibnamefont {Ulrich}}, \bibinfo
  {author} {\bibfnamefont {D.}~\bibnamefont {Richter}}, \bibinfo {author}
  {\bibfnamefont {M.}~\bibnamefont {Jetter}}, \bibinfo {author} {\bibfnamefont
  {P.}~\bibnamefont {Michler}}, \bibinfo {author} {\bibfnamefont
  {C.}~\bibnamefont {Roy}}, \ and\ \bibinfo {author} {\bibfnamefont
  {S.}~\bibnamefont {Hughes}},\ }\href {\doibase
  http://dx.doi.org/10.1103/PhysRevB.86.241304} {\bibfield  {journal} {\bibinfo
   {journal} {Phys. Rev. B}\ }\textbf {\bibinfo {volume} {86}},\ \bibinfo
  {pages} {241304} (\bibinfo {year} {2012})}\BibitemShut {NoStop}%
\bibitem [{\citenamefont {Ulhaq}\ \emph {et~al.}(2013)\citenamefont {Ulhaq},
  \citenamefont {Weiler}, \citenamefont {Roy}, \citenamefont {Ulrich},
  \citenamefont {Jetter}, \citenamefont {Hughes},\ and\ \citenamefont
  {Michler}}]{Ulhaq2013}%
  \BibitemOpen
  \bibfield  {author} {\bibinfo {author} {\bibfnamefont {A.}~\bibnamefont
  {Ulhaq}}, \bibinfo {author} {\bibfnamefont {S.}~\bibnamefont {Weiler}},
  \bibinfo {author} {\bibfnamefont {C.}~\bibnamefont {Roy}}, \bibinfo {author}
  {\bibfnamefont {S.}~\bibnamefont {Ulrich}}, \bibinfo {author} {\bibfnamefont
  {M.}~\bibnamefont {Jetter}}, \bibinfo {author} {\bibfnamefont
  {S.}~\bibnamefont {Hughes}}, \ and\ \bibinfo {author} {\bibfnamefont
  {P.}~\bibnamefont {Michler}},\ }\href {\doibase
  http://dx.doi.org/10.1364/OE.21.004382} {\bibfield  {journal} {\bibinfo
  {journal} {Opt. Express}\ }\textbf {\bibinfo {volume} {21}},\ \bibinfo
  {pages} {4382} (\bibinfo {year} {2013})}\BibitemShut {NoStop}%
\bibitem [{\citenamefont {Roszak}\ \emph {et~al.}(2005)\citenamefont {Roszak},
  \citenamefont {Grodecka}, \citenamefont {Machnikowski},\ and\ \citenamefont
  {Kuhn}}]{Roszak2005}%
  \BibitemOpen
  \bibfield  {author} {\bibinfo {author} {\bibfnamefont {K.}~\bibnamefont
  {Roszak}}, \bibinfo {author} {\bibfnamefont {A.}~\bibnamefont {Grodecka}},
  \bibinfo {author} {\bibfnamefont {P.}~\bibnamefont {Machnikowski}}, \ and\
  \bibinfo {author} {\bibfnamefont {T.}~\bibnamefont {Kuhn}},\ }\href {\doibase
  http://dx.doi.org/10.1103/PhysRevB.71.195333} {\bibfield  {journal} {\bibinfo
   {journal} {Phys. Rev. B}\ }\textbf {\bibinfo {volume} {71}},\ \bibinfo
  {pages} {195333} (\bibinfo {year} {2005})}\BibitemShut {NoStop}%
\bibitem [{\citenamefont {Machnikowski}\ and\ \citenamefont
  {Jacak}(2004)}]{machnikowski2004}%
  \BibitemOpen
  \bibfield  {author} {\bibinfo {author} {\bibfnamefont {P.}~\bibnamefont
  {Machnikowski}}\ and\ \bibinfo {author} {\bibfnamefont {L.}~\bibnamefont
  {Jacak}},\ }\href@noop {} {\bibfield  {journal} {\bibinfo  {journal} {Phys.
  Rev. B}\ }\textbf {\bibinfo {volume} {69}},\ \bibinfo {pages} {193302}
  (\bibinfo {year} {2004})}\BibitemShut {NoStop}%
\bibitem [{\citenamefont {Wilner}\ \emph {et~al.}(2014)\citenamefont {Wilner},
  \citenamefont {Wang}, \citenamefont {Thoss},\ and\ \citenamefont
  {Rabani}}]{Wilner2014}%
  \BibitemOpen
  \bibfield  {author} {\bibinfo {author} {\bibfnamefont {Y.~E.}\ \bibnamefont
  {Wilner}}, \bibinfo {author} {\bibfnamefont {H.}~\bibnamefont {Wang}},
  \bibinfo {author} {\bibfnamefont {M.}~\bibnamefont {Thoss}}, \ and\ \bibinfo
  {author} {\bibfnamefont {E.}~\bibnamefont {Rabani}},\ }\href@noop {}
  {\bibfield  {journal} {\bibinfo  {journal} {Phys. Rev. B}\ }\textbf {\bibinfo
  {volume} {89}},\ \bibinfo {pages} {205129} (\bibinfo {year}
  {2014})}\BibitemShut {NoStop}%
\bibitem [{\citenamefont {Weiss}\ \emph {et~al.}(2013)\citenamefont {Weiss},
  \citenamefont {Hutzen}, \citenamefont {Becker}, \citenamefont {Egger},\ and\
  \citenamefont {Thorwart}}]{Weiss2013}%
  \BibitemOpen
  \bibfield  {author} {\bibinfo {author} {\bibfnamefont {S.}~\bibnamefont
  {Weiss}}, \bibinfo {author} {\bibfnamefont {R.}~\bibnamefont {Hutzen}},
  \bibinfo {author} {\bibfnamefont {D.}~\bibnamefont {Becker}}, \bibinfo
  {author} {\bibfnamefont {R.}~\bibnamefont {Egger}}, \ and\ \bibinfo {author}
  {\bibfnamefont {M.}~\bibnamefont {Thorwart}},\ }\href@noop {} {\bibfield
  {journal} {\bibinfo  {journal} {Physica status solid}\ }\textbf {\bibinfo
  {volume} {250}},\ \bibinfo {pages} {2298} (\bibinfo {year}
  {2013})}\BibitemShut {NoStop}%
\bibitem [{\citenamefont {{B. A. Mason}}\ and\ \citenamefont
  {Hess}(1989)}]{Mason1989}%
  \BibitemOpen
  \bibfield  {author} {\bibinfo {author} {\bibnamefont {{B. A. Mason}}}\ and\
  \bibinfo {author} {\bibfnamefont {K.}~\bibnamefont {Hess}},\ }\href@noop {}
  {\bibfield  {journal} {\bibinfo  {journal} {Phys. Rev. B}\ }\textbf {\bibinfo
  {volume} {39}},\ \bibinfo {pages} {5051} (\bibinfo {year}
  {1989})}\BibitemShut {NoStop}%
\bibitem [{\citenamefont {Hughes}\ \emph
  {et~al.}(2009{\natexlab{a}})\citenamefont {Hughes}, \citenamefont {Christ},\
  and\ \citenamefont {Burghardt}}]{Hughes2009a}%
  \BibitemOpen
  \bibfield  {author} {\bibinfo {author} {\bibfnamefont {K.~H.}\ \bibnamefont
  {Hughes}}, \bibinfo {author} {\bibfnamefont {C.~D.}\ \bibnamefont {Christ}},
  \ and\ \bibinfo {author} {\bibfnamefont {I.}~\bibnamefont {Burghardt}},\
  }\href@noop {} {\bibfield  {journal} {\bibinfo  {journal} {J. Chem. Phys}\
  }\textbf {\bibinfo {volume} {131}},\ \bibinfo {pages} {024109} (\bibinfo
  {year} {2009}{\natexlab{a}})}\BibitemShut {NoStop}%
\bibitem [{\citenamefont {Cederbaum}\ \emph {et~al.}(2005)\citenamefont
  {Cederbaum}, \citenamefont {Gindensperger},\ and\ \citenamefont
  {Burghardt}}]{Burghardt2005}%
  \BibitemOpen
  \bibfield  {author} {\bibinfo {author} {\bibfnamefont {L.}~\bibnamefont
  {Cederbaum}}, \bibinfo {author} {\bibfnamefont {E.}~\bibnamefont
  {Gindensperger}}, \ and\ \bibinfo {author} {\bibfnamefont {I.}~\bibnamefont
  {Burghardt}},\ }\href {\doibase
  http://dx.doi.org/10.1103/PhysRevLett.94.113003} {\bibfield  {journal}
  {\bibinfo  {journal} {Phys. Rev. Lett.}\ }\textbf {\bibinfo {volume} {94}},\
  \bibinfo {pages} {113003} (\bibinfo {year} {2005})}\BibitemShut {NoStop}%
\bibitem [{\citenamefont {Gindensperger}\ \emph
  {et~al.}(2006{\natexlab{a}})\citenamefont {Gindensperger}, \citenamefont
  {Burghardt},\ and\ \citenamefont {Cederbaum}}]{Burghardt2006a}%
  \BibitemOpen
  \bibfield  {author} {\bibinfo {author} {\bibfnamefont {E.}~\bibnamefont
  {Gindensperger}}, \bibinfo {author} {\bibfnamefont {I.}~\bibnamefont
  {Burghardt}}, \ and\ \bibinfo {author} {\bibfnamefont {L.}~\bibnamefont
  {Cederbaum}},\ }\href {\doibase http://dx.doi.org/10.1063/1.2183304}
  {\bibfield  {journal} {\bibinfo  {journal} {J. Chem. Phys.}\ }\textbf
  {\bibinfo {volume} {124}},\ \bibinfo {pages} {144103} (\bibinfo {year}
  {2006}{\natexlab{a}})}\BibitemShut {NoStop}%
\bibitem [{\citenamefont {Gindensperger}\ \emph
  {et~al.}(2006{\natexlab{b}})\citenamefont {Gindensperger}, \citenamefont
  {Burghardt},\ and\ \citenamefont {Cederbaum}}]{Burghardt2006b}%
  \BibitemOpen
  \bibfield  {author} {\bibinfo {author} {\bibfnamefont {E.}~\bibnamefont
  {Gindensperger}}, \bibinfo {author} {\bibfnamefont {I.}~\bibnamefont
  {Burghardt}}, \ and\ \bibinfo {author} {\bibfnamefont {L.}~\bibnamefont
  {Cederbaum}},\ }\href {\doibase http://dx.doi.org/10.1063/1.2183305}
  {\bibfield  {journal} {\bibinfo  {journal} {J. Chem. Phys.}\ }\textbf
  {\bibinfo {volume} {124}},\ \bibinfo {pages} {144104} (\bibinfo {year}
  {2006}{\natexlab{b}})}\BibitemShut {NoStop}%
\bibitem [{\citenamefont {Tamura}\ \emph {et~al.}(2007)\citenamefont {Tamura},
  \citenamefont {Bittner},\ and\ \citenamefont {Burghardt}}]{Burghardt2007}%
  \BibitemOpen
  \bibfield  {author} {\bibinfo {author} {\bibfnamefont {H.}~\bibnamefont
  {Tamura}}, \bibinfo {author} {\bibfnamefont {E.}~\bibnamefont {Bittner}}, \
  and\ \bibinfo {author} {\bibfnamefont {I.}~\bibnamefont {Burghardt}},\ }\href
  {\doibase http://dx.doi.org/10.1063/1.2748050} {\bibfield  {journal}
  {\bibinfo  {journal} {J. Chem. Phys.}\ }\textbf {\bibinfo {volume} {127}},\
  \bibinfo {pages} {034706} (\bibinfo {year} {2007})}\BibitemShut {NoStop}%
\bibitem [{\citenamefont {Gindensperger}\ \emph {et~al.}(2007)\citenamefont
  {Gindensperger}, \citenamefont {Kappel},\ and\ \citenamefont
  {Cederbaum}}]{Gindensperger2007}%
  \BibitemOpen
  \bibfield  {author} {\bibinfo {author} {\bibfnamefont {E.}~\bibnamefont
  {Gindensperger}}, \bibinfo {author} {\bibfnamefont {H.}~\bibnamefont
  {Kappel}}, \ and\ \bibinfo {author} {\bibfnamefont {L.~S.}\ \bibnamefont
  {Cederbaum}},\ }\href {\doibase http://dx.doi.org/10.1063/1.2426342}
  {\bibfield  {journal} {\bibinfo  {journal} {J. Chem. Phys.}\ }\textbf
  {\bibinfo {volume} {126}},\ \bibinfo {pages} {034106} (\bibinfo {year}
  {2007})}\BibitemShut {NoStop}%
\bibitem [{\citenamefont {Tamura}\ \emph
  {et~al.}(2008{\natexlab{a}})\citenamefont {Tamura}, \citenamefont {Ramon},
  \citenamefont {Bittner},\ and\ \citenamefont {Burghardt}}]{Burghardt2008a}%
  \BibitemOpen
  \bibfield  {author} {\bibinfo {author} {\bibfnamefont {H.}~\bibnamefont
  {Tamura}}, \bibinfo {author} {\bibfnamefont {J.}~\bibnamefont {Ramon}},
  \bibinfo {author} {\bibfnamefont {E.}~\bibnamefont {Bittner}}, \ and\
  \bibinfo {author} {\bibfnamefont {I.}~\bibnamefont {Burghardt}},\ }\href
  {\doibase http://dx.doi.org/10.1103/PhysRevLett.100.107402} {\bibfield
  {journal} {\bibinfo  {journal} {Phys. Rev. Lett.}\ }\textbf {\bibinfo
  {volume} {100}},\ \bibinfo {pages} {107402} (\bibinfo {year}
  {2008}{\natexlab{a}})}\BibitemShut {NoStop}%
\bibitem [{\citenamefont {Tamura}\ \emph
  {et~al.}(2008{\natexlab{b}})\citenamefont {Tamura}, \citenamefont {Ramon},
  \citenamefont {Bittner},\ and\ \citenamefont {Burghardt}}]{Burghardt2008b}%
  \BibitemOpen
  \bibfield  {author} {\bibinfo {author} {\bibfnamefont {H.}~\bibnamefont
  {Tamura}}, \bibinfo {author} {\bibfnamefont {J.}~\bibnamefont {Ramon}},
  \bibinfo {author} {\bibfnamefont {E.}~\bibnamefont {Bittner}}, \ and\
  \bibinfo {author} {\bibfnamefont {I.}~\bibnamefont {Burghardt}},\ }\href
  {\doibase http://dx.doi.org/10.1021/jp077270p} {\bibfield  {journal}
  {\bibinfo  {journal} {J. Chem. Phys.}\ }\textbf {\bibinfo {volume} {B 112}},\
  \bibinfo {pages} {495} (\bibinfo {year} {2008}{\natexlab{b}})}\BibitemShut
  {NoStop}%
\bibitem [{\citenamefont {Hughes}\ \emph
  {et~al.}(2009{\natexlab{b}})\citenamefont {Hughes}, \citenamefont {Christ},\
  and\ \citenamefont {Burghardt}}]{Hughes2009b}%
  \BibitemOpen
  \bibfield  {author} {\bibinfo {author} {\bibfnamefont {K.~H.}\ \bibnamefont
  {Hughes}}, \bibinfo {author} {\bibfnamefont {C.~D.}\ \bibnamefont {Christ}},
  \ and\ \bibinfo {author} {\bibfnamefont {I.}~\bibnamefont {Burghardt}},\
  }\href {\doibase http://dx.doi.org/10.1063/1.3226343} {\bibfield  {journal}
  {\bibinfo  {journal} {J. Chem. Phys.}\ }\textbf {\bibinfo {volume} {131}},\
  \bibinfo {pages} {124108} (\bibinfo {year} {2009}{\natexlab{b}})}\BibitemShut
  {NoStop}%
\bibitem [{\citenamefont {Gindensperger}\ \emph
  {et~al.}(2006{\natexlab{c}})\citenamefont {Gindensperger}, \citenamefont
  {Burghardt},\ and\ \citenamefont {Cederbaum}}]{Gindensperger2006}%
  \BibitemOpen
  \bibfield  {author} {\bibinfo {author} {\bibfnamefont {E.}~\bibnamefont
  {Gindensperger}}, \bibinfo {author} {\bibfnamefont {I.}~\bibnamefont
  {Burghardt}}, \ and\ \bibinfo {author} {\bibfnamefont {L.}~\bibnamefont
  {Cederbaum}},\ }\href {\doibase http://dx.doi.org/10.1063/1.2183304}
  {\bibfield  {journal} {\bibinfo  {journal} {J. Chem. Phys.}\ }\textbf
  {\bibinfo {volume} {124}},\ \bibinfo {pages} {144103} (\bibinfo {year}
  {2006}{\natexlab{c}})}\BibitemShut {NoStop}%
\bibitem [{\citenamefont {Baugh}\ \emph {et~al.}(2005)\citenamefont {Baugh},
  \citenamefont {Moussa}, \citenamefont {Ryan}, \citenamefont {Nayak},\ and\
  \citenamefont {Laflamme}}]{Baugh2005}%
  \BibitemOpen
  \bibfield  {author} {\bibinfo {author} {\bibfnamefont {J.}~\bibnamefont
  {Baugh}}, \bibinfo {author} {\bibfnamefont {O.}~\bibnamefont {Moussa}},
  \bibinfo {author} {\bibfnamefont {C.~A.}\ \bibnamefont {Ryan}}, \bibinfo
  {author} {\bibfnamefont {A.}~\bibnamefont {Nayak}}, \ and\ \bibinfo {author}
  {\bibfnamefont {R.}~\bibnamefont {Laflamme}},\ }\href {\doibase
  http://dx.doi.org/10.1038/nature04272} {\bibfield  {journal} {\bibinfo
  {journal} {Nature}\ }\textbf {\bibinfo {volume} {438}},\ \bibinfo {pages}
  {470} (\bibinfo {year} {2005})}\BibitemShut {NoStop}%
\bibitem [{\citenamefont {Taylor}\ \emph {et~al.}(2003)\citenamefont {Taylor},
  \citenamefont {Marcus},\ and\ \citenamefont {Lukin}}]{Taylor2003}%
  \BibitemOpen
  \bibfield  {author} {\bibinfo {author} {\bibfnamefont {J.~M.}\ \bibnamefont
  {Taylor}}, \bibinfo {author} {\bibfnamefont {C.~M.}\ \bibnamefont {Marcus}},
  \ and\ \bibinfo {author} {\bibfnamefont {M.~D.}\ \bibnamefont {Lukin}},\
  }\href {\doibase http://dx.doi.org/10.1103/PhysRevLett.90.206803} {\bibfield
  {journal} {\bibinfo  {journal} {Phys. Rev. Lett.}\ }\textbf {\bibinfo
  {volume} {90}},\ \bibinfo {pages} {206803} (\bibinfo {year}
  {2003})}\BibitemShut {NoStop}%
\bibitem [{\citenamefont {Wang}\ \emph {et~al.}(2008)\citenamefont {Wang},
  \citenamefont {Pietz}, \citenamefont {Walowski}, \citenamefont {Frster},
  \citenamefont {Lepsa},\ and\ \citenamefont {Mnzenberg}}]{Wang2008}%
  \BibitemOpen
  \bibfield  {author} {\bibinfo {author} {\bibfnamefont {Z.}~\bibnamefont
  {Wang}}, \bibinfo {author} {\bibfnamefont {M.}~\bibnamefont {Pietz}},
  \bibinfo {author} {\bibfnamefont {J.}~\bibnamefont {Walowski}}, \bibinfo
  {author} {\bibfnamefont {A.}~\bibnamefont {Frster}}, \bibinfo {author}
  {\bibfnamefont {M.}~\bibnamefont {Lepsa}}, \ and\ \bibinfo {author}
  {\bibfnamefont {M.}~\bibnamefont {Mnzenberg}},\ }\href {\doibase
  http://dx.doi.org/10.1063/1.2940734} {\bibfield  {journal} {\bibinfo
  {journal} {J. Appl. Phys.}\ }\textbf {\bibinfo {volume} {103}},\ \bibinfo
  {pages} {123905} (\bibinfo {year} {2008})}\BibitemShut {NoStop}%
\end{thebibliography}%
\end{document}